\documentclass[prd,preprint,nofootinbib,showpacs]{revtex4}

\usepackage{epsf,epsfig,subfigure,graphicx,amsmath,amssymb}

\usepackage{amsfonts}
\usepackage{latexsym}
\usepackage{color}
\usepackage[normalem]{ulem}
\usepackage{accents}
\usepackage{tikz}

\newcommand{\dis}[1]{\begin{equation}\begin{split}#1\end{split}\end{equation}}
\newcommand{\be}{\begin{equation}}
\newcommand{\ee}{\end{equation}}
\def\bea{\begin{eqnarray}}
\def\eea{\end{eqnarray}}

\newcommand{\bfrac}[2]{{\left(\frac{#1}{#2} \right)  }}
\newcommand{\VEV}[1]{\langle #1 \rangle}

\newcommand{\Mp}{M_P}
\newcommand{\calP} {{\cal P}}

\newcommand\gev{\,{\rm GeV}}

\newcommand\mpc{\,{\rm Mpc}}

\newcommand\fnl{{f_{\rm NL}}}

\newcommand\dilaton{{\rm dilaton }}
\newcommand\bfk{{\bf k}}     
\newcommand\bfx{{\bf x}}     
\newcommand\bfp{{\bf p}}

\begin{document}

\title{\LARGE Primordial perturbations  \\ from dilaton-induced gauge fields}

\author{
Kiwoon Choi$\,^a$\footnote{kchoi@ibs.re.kr}, 
Ki-Young Choi$\,^b$\footnote{kiyoungchoi@kasi.re.kr}, 
Hyungjin Kim$\,^c$\footnote{hjkim06@kaist.ac.kr}, 
Chang Sub Shin$\,^d$\footnote{changsub@physics.rutgers.edu}
}

 \affiliation{$^a$Center for Theoretical Physics of the Universe, Institute for Basic Science (IBS), Daejeon 305-811, Korea
 \\
$^b$Korea Astronomy and Space Science Institute, Daejeon 305-348,  Republic of Korea 
\\
$^c$Department of Phyiscs, KAIST, Daejeon 305-701, Korea
\\
$^d$New High Energy Theory Center, Department of Physics and Astronomy, Rutgers University, Piscataway, NJ 08854, USA
}

\begin{abstract}

We study the primordial scalar and tensor  perturbations in  inflation scenario involving a spectator dilaton field. In our setup, the rolling spectator dilaton causes a tachyonic instability of gauge fields, leading to a copious production of gauge fields in the superhorizon regime, which generates additional scalar and tensor perturbations through  gravitational interactions. Our prime concern is the possibility to enhance the tensor-to-scalar ratio $r$ relative to the standard result, while satisfying the observational constraints. To this end, we allow  the dilaton field to be stabilized before the end of inflation, but after the CMB scales exit the horizon.  We show that for the inflaton slow roll parameter $\epsilon \gtrsim 10^{-3}$, the tensor-to-scalar ratio in our setup can be enhanced only by a factor of ${\cal O}(1)$ compared to the standard result. On the other hand, for smaller $\epsilon$ corresponding to a lower inflation energy scale, a much bigger enhancement can be achieved, so that our setup can give rise to an observably large $r\gtrsim 10^{-2}$ even when $\epsilon\ll 10^{-3}$. The tensor perturbation sourced by the spectator dilaton can have a strong scale dependence, and is generically red-tilted. We also discuss a specific model to realize our scenario, and identify the parameter region giving an observably large $r$ for relatively low inflation energy scales.

\end{abstract}


\preprint{CTPU-15-08}
\preprint{RUNHETC-2015-05}

\vspace*{3cm}
\maketitle
\tableofcontents


\section{Introduction}
\label{introduction}

The cosmological inflation not only solves  the naturalness problems in the standard big bang cosmology, but also provides an appealing mechanism to generate the seed of the large scale structure and the cosmic microwave background temperature anisotropies in the present universe \cite{review}. 
During the inflationary phase, primordial gravitational waves can be generated  from the quantum fluctuation of metric. The latest joint analysis  of  BICEP2/Keck Array and Planck data provides an upper limit on such tensor perturbation, implying that the tensor-to-scalar ratio is bounded as~\cite{Ade:2015tva},
\dis{
r\equiv \frac{{\cal P}_t }{{\cal P}_\zeta} \, <\, 0.12\quad {\rm (95\% \,CL)} \label{limitr}
}
at the pivot scale $k_*=0.05 \mpc^{-1}$.  In the minimal single field inflation scenario, this can be used to constrain the energy scale of inflation based on the standard relation between the tensor power spectrum and  the inflationary Hubble scale~\cite{Lyth:1996im}: 
\dis{
{\cal P}_t = \frac{2H^2}{\pi^2\Mp^2}.\label{Pt}
}

On the other hand, many of the well motivated models of particle physics involve a light scalar field  which couples to gauge fields in a way to provide  an additional source of perturbations.
For instance, if the scalar field evolves appropriately during the inflationary epoch, it can cause a tachyonic instability of gauge fields, leading to a copious production of gauge fields. Then the produced gauge fields may result in a significant amount of additional tensor perturbations, so modify the  relation (\ref{Pt})~\cite{Senatore:2011sp,Fujita:2014oba,Biagetti:2014asa,Mirbabayi:2014jqa, Ozsoy:2014sba}. A well studied example is an axion-like field $\eta$ which couples to gauge field as~\cite{Pajer:2013fsa,Anber:2009ua,Barnaby:2010vf,Barnaby:2011vw,Barnaby:2012xt,Barnaby:2011qe,Ferreira:2014zia,Adshead:2015pva,Eccles:2015ipa}
\dis{
\label{axion-coupling}
\Delta {\cal L}_{\rm axion} \,=\, \frac{1}{32\pi^2} \frac{\eta}{f} F_{\mu\nu} \tilde{F}^{\mu\nu},
}
where $F_{\mu\nu} = \partial_\mu A_\nu -\partial_\nu A_\mu$ is the gauge field strength, $\tilde{F}_{\mu\nu}= \frac{1}{2} \epsilon_{\mu\nu\rho\sigma} F^{\rho\sigma}$ is its dual, and $f$ is the axion decay constant. Regardless of whether it is an inflaton or just a spectator field, a rolling axion  with the coupling (\ref{axion-coupling}) can generate additional tensor modes which are highly non-gaussian~\cite{Cook:2013xea}, parity-violating~\cite{Saito:2007kt,Sorbo:2011rz,Cook:2011hg, Crowder:2012ik,Shiraishi:2013kxa, Bartolo:2014hwa}, and blue-tilted~\cite{Mukohyama:2014gba}. However, if the coupling (\ref{axion-coupling}) is strong enough to enhance the tensor-to-scalar ratio significantly, it can lead a large non-gaussianity 
in scalar perturbation, which is in danger to be incompatible with the recent Planck results~\cite{Ade:2015ava}.

There is another type of well motivated light scalar field, a \dilaton $\sigma$ (or moduli field) which couples to gauge fields as
\dis{
\label{dilaton-coupling}
\Delta {\cal L}_{\rm dilaton}\, =\, -\frac{I^2(\sigma)}{4} F_{\mu\nu}F^{\mu\nu},
}
where $g=I^{-1}(\sigma)$ can be identified as  the field-dependent gauge coupling. As in the case of axion, a rolling dilaton can produce gauge fields by causing a tachyonic instability. Cosmological implications of  rolling dilaton with the coupling (\ref{dilaton-coupling}) have been studied extensively in the context of {inflationary magnetogenesis}~\cite{Ratra:1991bn,Bamba:2003av,Martin:2007ue,Kobayashi:2014sga}. As the produced gauge fields are stretched out the horizon during inflation, it may provide the origin of large scale magnetic fields in the present universe. However this mechanism of magnetogenesis is constrained in several ways. Requiring that the electromagnetic energy density should not exceed the inflaton energy density, either the amplitude of the produced magnetic field should be too small to explain the large scale magnetic field in the present universe~\cite{Kanno:2009ei,Demozzi:2009fu,Fujita:2012rb,Ferreira:2014mwa}, or the gauge coupling $g=I^{-1}$ should run from an extremely large value to ${\cal O}(1)$~\cite{Demozzi:2009fu}. In addition, the produced electromagnetic field contributes to the primordial density perturbations, providing a variety of additional constraints on this mechanism of magnetogenesis~\cite{Bonvin:2011dt,Suyama:2012wh,Giovannini:2013rme,Ringeval:2013hfa,Fujita:2013qxa,Fujita:2014sna, Caldwell:2011ra,Barnaby:2012tk,Motta:2012rn,Jain:2012ga,Jain:2012vm,Nurmi:2013gpa, Bartolo:2012sd,Thorsrud:2013kya, Watanabe:2009ct, Soda:2012zm,Deskins:2013lfx}.

In this paper, we study systematically the scalar and tensor perturbations sourced by a rolling spectator dilaton which couples to gauge field kinetic terms as (\ref{dilaton-coupling}), while taking into account the known observational constraints. Our prime concern is the possibility to enhance the tensor-to-scalar ratio $r$ relative to the standard result $r=16\epsilon$ of the  single field inflation scenario, where $\epsilon$ is the inflaton slow roll parameter.  
To this end, we allow the spectator  \dilaton  to be stabilized before the end of inflation, but after the CMB scales exit the horizon. As we will see,
this makes it possible that $r$ is large enough to be observable in the near future, e.g. $r\gtrsim 10^{-2}$, even when the inflation energy scale is relatively low to give $\epsilon\ll 10^{-3}$.

Although our scheme reduces to the conventional single field inflation after the dilaton is stabilized, the dilaton
dynamics which took place before the stabilization leaves an imprint on the primordial power spectrum that exit the horizon while the dilaton field is rolling. Imposing the known observational constraints on the scalar perturbation sourced by rolling dilaton, we find that for the inflaton slow roll parameter $\epsilon  \gtrsim 10^{-3}$, the tensor-to-scalar ratio $r$ can be modified  only by a factor of ${\cal O}(1)$ compared to the standard result. However, for smaller $\epsilon$, which corresponds to a lower inflation energy scale, $r$ can be enhanced by a much larger factor. Specifically, the tensor perturbation sourced by rolling dilaton can be large enough to give $r\gtrsim 10^{-2}$, while satisfying the observational constraints, even when the inflaton slow roll parameter $\epsilon\ll 10^{-3}$. We also find that the tensor power spectrum in this case can have a strong scale dependence, which is generically red-tilted.

Compared to the axion case, our dilaton scenario has several distinctive features. For instance, both polarizations of tensor mode are equally produced in the dilaton case, while only a certain polarization state is produced in the axion case.  {Another difference is in the scale dependence. In case that the axion or dilaton coupling to gauge fields is strong enough to generate a large tensor perturbation, the energy density of dilaton-induced gauge fields continues to be growing over the superhorizon regime, while the energy density of axion-induced gauge fields is diluted soon after the horizon crossing. As a result, for the dilaton case the perturbations are produced dominantly in the superhorizon regime, while for the axion case the production of perturbations is active only around the horizon crossing. This results in a strongly red-tilted tensor spectral index for the dilaton case, which is not suppressed by slow roll parameters. On the other hand, for the axion case the tensor spectral index is suppressed by slow roll parameters, although it can be numerically sizable and blue-tilted~\cite{Mukohyama:2014gba}.

This paper is organized as follows. We describe our setup in Section~\ref{setup}, and compute the resulting scalar and tensor perturbations in  Section~\ref{scalarpert} and~\ref{tensorpert}, respectively. In Section~\ref{model}, we discuss the implications of our result and present a specific model with interesting observational consequences. Section~\ref{conclusion} is the conclusion.

\section{Setup} \label{setup}

We consider an inflationary cosmology described by 
\begin{eqnarray}
	S &=& \int d^4 x \sqrt{-g} \left[ \frac{\Mp^2}{2} R + {\cal
	L}_{\rm inf}(\phi) - \frac{1}{2}\partial_\mu \sigma \partial^\mu \sigma - V(\sigma)
	- \frac{I^2(\sigma)}{4} F_{\mu\nu} F^{\mu\nu}\right] ,
	\label{Model}
\end{eqnarray}
where $\Mp \simeq 2.4\times 10^{18}$ GeV is the reduced Planck mass,  ${\cal L}_{\rm inf}(\phi)$ is the lagrangian density of the inflaton field $\phi$,
and $F_{\mu\nu}$ is the field strength of $U(1)$ gauge field which couples to the  dilaton field $\sigma$. For simplicity, here we assume that there is no direct coupling of the inflaton to the  dilaton and gauge fields. We assume also that the inflaton field $\phi$ satisfies the conventional slow-roll conditions, and the total energy density is dominated by the inflaton energy density over the whole period of inflation. We will use the spacially flat gauge, for which the metric perturbations are parametrized as
\begin{eqnarray}
	ds^2 = a^2(\tau) \left[ -(1+2\Phi) d\tau^2 + 2 \partial_i
	B d\tau dx^i + (\delta_{ij}+h_{ij}) dx^i dx^j\right],
	\label{metric}
\end{eqnarray}
where the conformal time coordinate $\tau$ is defined as $d\tau = dt/a(t)$ for the Robertson-Walker time coordinate $t$,
and $h_{ij}$ satisfies the traceless/transverse condition, $ h_{ii} =\partial_i h_{ij} =0$. The inflaton, dilaton, and gauge field are expanded also around a homogeneous background as 
\begin{eqnarray}
	\phi (\tau,\bfx) &=&  \phi_0(\tau) +  \delta \phi(\tau,\bfx), \nonumber
	\\
	\sigma (\tau,\bfx) &=&  \sigma_0(\tau) +  \delta \sigma(\tau,\bfx), \nonumber 
	\\
	A_\mu (\tau,\bfx) &=&  \delta A_\mu(\tau,\bfx). 
\end{eqnarray}

\subsection{Gauge field production by a rolling dilaton}

As is well known, a rolling dilaton field during inflation can develop a tachyonic instability of gauge field, leading  to a copious production of gauge fields in the superhorizon regime. Choosing the gauge condition $\nabla \cdot \vec{A} = 0$ and $A_0=0$, the equation of motion of gauge field is given by
\bea
\left( \partial_\tau^2 + 2 \frac{\partial_\tau I}{I} \partial_\tau - \nabla^2 \right) A_i(\tau,\bfx) = 0.
\eea
After the Fourier expansion 
\bea
A_i(\tau, \bfx) &=& 
\int \frac{d^3\bfk}{(2\pi)^{3/2}} \hat{A}_i(\tau,\bfk) e^{i \bfk \cdot \bfx}, \nonumber \\
\hat{A}_i(\tau,\bfk) &=& \sum_{\lambda=\pm} \epsilon_{i,\lambda} (\hat \bfk)
[A_\lambda(\tau,k) a_{\bfk, \lambda} 
+ A^*_\lambda(\tau,k) a^\dagger_{-\bfk,\lambda} ],\nonumber
\eea
it is convenient to redefine the gauge field mode function as \bea
V_\lambda \equiv I(\sigma_0) A_\lambda,\eea
where $I(\sigma_0)$ depends only on the background dilaton field $\sigma_0(\tau)$, not on the dilaton fluctuation.
Then the equation of motion of the canonically normalized mode function is given by
\begin{eqnarray}
	\partial_\tau^2V_{\lambda} (\tau,k) 
	+ \left( k^2 - \frac{\partial_{\tau}^2I}{I} \right) V_{\lambda}(\tau, k) &=& 0.
	\label{eomV}
\end{eqnarray}
Note that both helicity states evolve in the same way, so we can drop the helicity index from now. This is different from the axion case where different helicity state experiences different evolution, which results in  parity violating phenomena.

The details of gauge field production by rolling dilaton depends on the functional form of the dilaton coupling $I(\sigma)$.  For canonically normalized dilaton field, a particularly well motivated form of the dilaton coupling  is 
\bea I(\sigma) = e^{\sigma/\Lambda},\eea 
where $\Lambda$ is a constant mass parameter. In this case, the evolution rate of the dilaton coupling (relative to the Hubble expansion rate) is given by 
\bea
n\,\equiv\, -\frac{\dot{I}}{HI} \,=\, -\frac{\dot{\sigma}}{H\Lambda}. \label{evolution-rate} \eea
If the spectator dilaton underwent a time evolution satisfying  
\bea 
\label{slow-dilaton} |\ddot{\sigma}|\ll H|\dot{\sigma}|,  \eea
where the dot denotes the derivative with respect to the Robertson-Walker time coordinate $t$,
one finds
\bea
\left|\frac{\dot{n}}{Hn}\right| \,=\, \left|\frac{\dot{H}}{H^2} -\frac{\ddot{\sigma}}{H\dot{\sigma}}\right|\,\ll\, 1. \eea
This suggests that the evolution rate $n$ can be approximated as a constant over a certain duration of the dilaton rolling.

To examine the possibility to enhance the tensor-to-scalar ratio $r$, in this paper we consider a scenario that the spectator dilaton rolls over a period of the e-folding number   $\Delta N ={\cal O}(10)$, under the assumption that both the inflaton and the dilaton began to roll at a similar time. Then, as long as  $|\ddot{\sigma}|/H|\dot{\sigma}| \lesssim \mbox{few}\times 10^{-2}$, the evolution rate $n$ can be approximated as a constant over the entire period of the dilaton rolling. Note that in our scenario, the dilaton field is stabilized before the end of inflation, and therefore $\Delta N$ can be significantly smaller than the total e-folding number $N_T \gtrsim 50-60$ of inflation. For simplicity, we  assume that the transition from the rolling dilaton phase to the stabilized dilaton phase  takes place within a short time interval $\Delta t \ll 1/H$. Then the dilaton-dependent gauge coupling  evolves as 
\begin{eqnarray}
\label{dilaton-evol}
I(\tau) \,\equiv\, I(\sigma_0(\tau)) \propto a(\tau)^{-n},
\end{eqnarray}
where $n$ is a nonzero constant during the rolling phase, but $n=0$ right after the dilaton is stabilized. 
This might be a rather crude approximation for the real dilaton dynamics, but is sufficient for our purpose to explore the possibility to enhance $r$. The reason to consider a dilaton field stabilized before the end of inflation is that it allows $r$ to be enhanced by a large factor  while satisfying the observational constraints. If the dilaton field  rolls until the end of inflation, whenever $r$ is significantly affected, scalar perturbation is dominated by the contribution from the rolling dilaton, which would lead to a too large deviation of the scalar spectral index from the observed value, or a too large non-gaussianity.

The evolution rate $n$ in (\ref{evolution-rate}) can be either positive or negative. Note that changing the sign of $n$ amounts to $g\rightarrow g^{-1}$ for the gauge coupling $g$. For a positive $n$, the field-dependent gauge coupling $g= I^{-1}$ runs from the weak coupling regime to the strong coupling regime. For simplicity, we will focus on the case of positive $n$ with $g\lesssim 1$, where the production of electric fields dominates over the production of magnetic fields. This choice of $n$ opens a possibility that the $U(1)$ gauge field in our setup can be identified as one of the standard model gauge fields if $I(\sigma)={\cal O}(1)$ after the dilaton is stabilized.

For the dilaton coupling (\ref{dilaton-evol}), the equation of motion of the gauge field mode takes the form
\begin{eqnarray}
	\partial_\tau^2V + \left[ k^2 - \frac{n(n-1)}{\tau^2} \right] V = 0.
\end{eqnarray}
Imposing the Bunch-Davies initial condition,
$$
\lim_{k\tau \rightarrow -\infty}V(\tau,k) = \frac{e^{-i k \tau}}{\sqrt{2k}},
$$ 
the solution is given by
\begin{eqnarray}
	V(\tau, k) = \frac{1}{\sqrt{2k}} \sqrt{\frac{-k\tau\pi}{2}} 
	H^{(1)}_{n-1/2}(-k\tau),
\end{eqnarray}
where $H_\nu^{(1)}$ is the Hankel function of the first kind.
Using the asymptotic form of the Hankel function:
\dis{
H^{(1)}_\nu (z) \simeq 
-\frac{i \Gamma(\nu)}{\pi} \left( \frac{2}{z}\right)^\nu
+ \frac{1}{\Gamma(\nu+1)} \left(\frac{z}{2} \right)^\nu 
- \frac{i\Gamma(-\nu)}{\pi} \cos \nu \pi \left(\frac{z}{2}\right)^\nu
\quad{\rm for} \quad z\ll1 ,
\label{Hasym}
}
we find that
the gauge field mode in the superhorizon regime with $|k\tau| \ll1$ is given by
\begin{eqnarray}
	V(\tau, k)\, \simeq\, -\frac{i}{\sqrt{2k}}
	\frac{\Gamma(n-1/2)}{\sqrt{\pi}} \left( \frac{2}{-k\tau} \right)^{n-1},
\end{eqnarray}
where the blow up of the amplitude in the superhorizon  limit $| k\tau| \rightarrow 0$ (for $n>1$) is due to the tachyonic instability of gauge field caused by the rolling dilaton.

For subsequent discussion, it is convenient to define the electric and magnetic fields as
\dis{
{E}_i(\tau,{\bf x}) = -\frac{I}{a^2} \partial_\tau{A}_i (\tau,{\bf x}),\qquad 
{B}_i(\tau,{\bf x}) =  \frac{I}{a^2} (\nabla \times \vec{A})_i,
\label{EBA}
}
for which the energy density of the $U(1)$ gauge field is given by  
\dis{
\rho_{U(1)} \equiv T^{U(1)}_{tt}= \frac{1}{2}(|\vec{E}|^2+|\vec{B}|^2).
\label{rhoem}
} 
One can now make the Fourier expansion:
\bea
E_i(\tau,\bfx) &=& 
\int\frac{d^3\bfk}{(2\pi)^{3/2}} \widehat{E}_i(\tau,\bfk) e^{i \bfk\cdot\bfx}, \nonumber \\
\widehat{E}_i(\tau,\bfk) 
&=& \sum_{\lambda} \epsilon_{i,\lambda}(\hat \bfk)
\left[{\cal E}(\tau,k) a_{\bfk,\lambda} + {\cal E}^*(\tau,k) a_{-\bfk,\lambda}^\dagger\right], 
\nonumber\\
B_i(\tau,\bfx) &=&
\int\frac{d^3\bfk}{(2\pi)^{3/2}} \widehat{B}_i(\tau,\bfk) e^{i \bfk\cdot\bfx},
\nonumber \\
\widehat{B}_i(\tau,\bfk) 
&=& \sum_{\lambda} \lambda \epsilon_{i,\lambda}(\hat \bfk)
\left[{\cal B}(\tau,k) a_{\bfk,\lambda} + {\cal B}^*(\tau,k) a_{-\bfk,\lambda}^\dagger\right], \nonumber
\eea
where the corresponding electric and magnetic mode functions are given by
\bea
 {\cal E}(\tau,k) 
 &=&  -\frac{1}{a^2} \sqrt{\frac{k}{2}} \sqrt{\frac{-k\tau \pi}{2}} H^{(1)}_{n+1/2}(-k\tau) \nonumber \\
 &\simeq& \frac{i\Gamma(n+1/2)}{\sqrt{\pi}} \sqrt{\frac{k}{2}} (H\tau)^2 
 \left(\frac{2}{-k\tau}\right)^{n} \quad {\rm for}\quad |k\tau|\ll 1,
 \label{EMode}
 \\
 {\cal B}(\tau,k) 
 &=&\frac{1}{a^2} \sqrt{\frac{k}{2}} \sqrt{\frac{-k\tau \pi}{2}} 
 H^{(1)}_{n-1/2}(-k\tau) \nonumber \\
 &\simeq& -\frac{i\Gamma(n-1/2)}{\sqrt{\pi}} \sqrt{\frac{k}{2}} (H\tau)^2 
 \left(\frac{2}{-k\tau}\right)^{n-1} \quad {\rm for} \quad |k\tau|\ll 1.
  \label{BMode}
\eea
Note that the last approximation for ${\cal E}$ and ${\cal B}$  are valid only for $n \geq \frac{1}{2}$.
 Otherwise the latter two terms in (\ref{Hasym}) become important.
 Note also that the electric field always dominates over the magnetic field in the superhorizon regime with $|k\tau| \ll1$. For a given mode, the electric field on superhorizon scale  decreases ($n<2$), remains constant ($n=2$), and grows ($n>2$). 
 As we will see in the subsequent two sections,  the gauge fields produced by rolling dilaton can significantly affect the scalar and tensor perturbations
 when $n> 2$.

\section{Scalar perturbation} \label{scalarpert}

In the spacially flat gauge,  the curvature perturbation $\cal R$ is given by
\bea
{\cal R}  = - H \frac{\delta q}{\rho+p},
\eea
where $\delta q$ is the scalar 3-momentum potential defined as  $\partial_i\delta q=\delta T^t_i$ for the energy momentum tensor perturbation $\delta T^\mu_\nu$. In the multi-fluid case, it can be decomposed as
\dis{
{\cal R} = \sum_\alpha \frac{(\rho+p)_\alpha}{(\rho + p)} {\cal R}_\alpha\quad {\rm for}\quad
{\cal R}_\alpha \equiv  - H \frac{\delta q_\alpha}{(\rho + p)_\alpha},
}
where $\alpha$ denotes the  fluid species.  
In our scenario, the dilaton and gauge field fluctuations could constitute an important part of the total curvature perturbation during the rolling phase of dilaton. However, after the dilaton is stabilized, the dilaton perturbation becomes a massive field, and gauge fields are no longer produced. Then the dilaton and gauge field contributions to $\delta q$  are quickly diluted  away as  $\delta q_\sigma \propto a^{-3}$ and $\delta q_{A_\mu}\propto a^{-4}$. If the universe has experienced a sufficient inflationary expansion after the dilaton is stabilized, which is the case of our prime interest, the curvature perturbation at the end of inflation is determined simply by the inflaton perturbation as
\dis{
{\cal R} \, \simeq\,  {\cal R}_\phi \,=\, H\frac{\delta \phi }{\dot{\phi}}.\label{R_phi}
} 
In fact, if the dilaton keeps rolling until the end of inflation, whenever tensor perturbation is significantly affected, scalar perturbation is dominated by the contribution sourced by rolling dilaton. Such scenario then yields a too large spectral index and non-gaussianity to be compatible with the observational constraints~\cite{Suyama:2012wh}. In the following, we compute the inflaton perturbation at the end of inflation, including the effect of pre-evolution
during the period before the dilaton stabilization.

\subsection{Evolution of the inflaton and dilaton perturbations}
The equations of motion for the background inflaton and dilaton fields are given by
\begin{eqnarray}
	\phi_0'' + 2 {\cal H} \phi_0' + a^2 \partial_\phi V(\phi_0) &=& 0, \nonumber
	\\
	\sigma_0'' + 2 {\cal H} \sigma_0' + a^2 \partial_\sigma V(\sigma_0) &=& 
	a^2 \frac{\partial_\sigma I}{I} \VEV{|\vec{E}|^2 - |\vec{B}|^2},
\end{eqnarray}
where the prime denotes the derivative with respect to the conformal time coordinate $\tau$, and ${\cal H} \equiv a'/a$.
Assuming a slow-roll motion of the background fields, and also neglecting the back-reaction effects, we obtain the equations of motion of perturbations as 
\begin{eqnarray}
	\delta \phi '' + 2 {\cal H} \delta\phi' + k^2 \delta \phi
	+ a^2 \left( \partial_\phi^2V - 3 \frac{\dot{\phi}_0^2}{\Mp^2} \right) 
	\delta \phi
	- 3 a^2 \frac{\dot{\sigma}_0 \dot{\phi}_0}{\Mp^2} \delta \sigma
	&=&  S_1 (\tau,\bfk),
	\label{IPEOM}
	\\
	\delta \sigma '' + 2 {\cal H} \delta\sigma' + k^2 \delta \sigma
	+ a^2 
	\left( \partial_\sigma^2 V - 3 \frac{\dot{\sigma}_0^2}{\Mp^2} \right) 
	\delta \sigma
	- 3 a^2 \frac{\dot{\sigma}_0 \dot{\phi}_0}{\Mp^2} \delta \phi
	&=&  S_2 (\tau,\bfk) + S_3(\tau,\bfk),
	\label{DPEOM}
\end{eqnarray}
where the source terms $S_i$ ($i=1,2,3$) in the momentum space are given by
\bea
S_1(\tau,\bfk) &=& \frac{a^2 \dot{\phi}_0}{2\Mp^2 H} 
\int \frac{d^3\bfp}{(2\pi)^{3/2}} \frac{(k-p)_i p_j}{k^2}
\left[
\widehat{E}_i(\tau,{\bf p}) \widehat{E}_j(\tau,{\bfk - \bf p}) + 
\widehat{B}_i(\tau,{\bf p}) \widehat{B}_j(\tau,{\bfk - \bf p})
\right],
\label{S1} \nonumber
\\
S_2(\tau,\bfk) &=& a^2\frac{\partial_\sigma I}{I}
\int \frac{d^3\bfp}{(2\pi)^{3/2}} 
\left[
\widehat{E}_i(\tau,{\bf p}) \widehat{E}_i(\tau,{\bfk - \bf p}) + 
\widehat{B}_i(\tau,{\bf p}) \widehat{B}_i(\tau,{\bfk - \bf p})
\right],
\label{S2} \nonumber
\\
S_3(\tau,\bfk) &=& \frac{a^2 \dot{\sigma}_0}{2\Mp^2 H} 
\int \frac{d^3\bfp}{(2\pi)^{3/2}} \frac{(k-p)_i p_j}{k^2},
\left[
\widehat{E}_i(\tau,{\bf p}) \widehat{E}_j(\tau,{\bfk - \bf p}) + 
\widehat{B}_i(\tau,{\bf p}) \widehat{B}_j(\tau,{\bfk - \bf p})
\right].
\label{S3}\nonumber
\eea
See Appdenix.~\ref{SourceDerivation} for the derivation of the above equations of motion. The source terms  $S_1(\tau,\bfk)$ and $S_3(\tau,\bfk)$ are due to the gravitational interaction between the inflaton/dilaton fluctuation and the gauge fields produced by the rolling background dilaton, while $S_2(\tau,\bfk)$ originates  from the direct coupling between the \dilaton and gauge fields. As can be seen from (\ref{IPEOM}) and (\ref{DPEOM}), even though there is no direct coupling between the inflaton and dilaton, their perturbations can be mixed with each other by gravitational interaction. As a result, the inflaton perturbation can be significantly affected by the \dilaton perturbation sourced  by $S_2$ and $S_3$. As we will see later, the inflaton perturbation sourced by gauge fields comes dominantly from the source term  $S_2(\tau,\bfk)$.

Let us divide the inflaton perturbation $\delta\phi$ into four pieces,
\bea
\delta\phi = \delta\phi^{(v)} + \delta\phi^{(S_1)} + \delta\phi^{(S_2)} + \delta\phi^{(S_3)},
\eea
where $\delta\phi^{(v)}$ represents the piece from vacuum fluctuation, while $\delta\phi^{(S_i)}$  ($i=1,2,3$) represent the parts induced by the source terms $S_i$. To obtain the solution, it is convenient to rotate the field basis into the propagation eigenbasis. For this, we rewrite (\ref{IPEOM}) and~(\ref{DPEOM}) as~\cite{Byrnes:2006fr,Ferreira:2014zia}
\begin{eqnarray}
	\left[ 
		\partial_\tau^2 +  (k^2 - \frac{2}{\tau^2}) +  \frac{1}{\tau^2} 
		\left( 
		\begin{array}{cc}
			\Delta_\phi & \Delta  \\
			\Delta & \Delta_\sigma
		\end{array}
		\right)
	\right] 
	\left(
	\begin{array}{c}
		a \delta\phi \\
		a \delta\sigma
	\end{array}
	\right)
	=
	a(\tau)\left(
	\begin{array}{c}
		S_1 \\
		S_2 + S_3
	\end{array}
	\right),
\end{eqnarray}
where 
\dis{
\Delta_\alpha  & = \,\frac{\partial_\alpha^2 V - 3 \dot{\alpha}^2_0 / \Mp^2 }{H^{2}}-3\epsilon 
\,\simeq\, 3(\eta_{\alpha}-2\epsilon_\alpha)  -3\epsilon \quad(\alpha=\phi, \sigma),\\
\Delta &=\,  -\frac{3 \dot{\phi}_0\dot{\sigma}_0}{\Mp^2 H^2} \,\simeq\, -6\sqrt{\epsilon_\phi \epsilon_\sigma
}}
for the slow roll parameters
\bea
\epsilon_\alpha \,\equiv\, {\frac{\Mp^2}{2} \bfrac{\partial_\alpha V}{V}^2}, \quad 
\eta_\alpha\,\equiv \Mp^2 \left(\frac{\partial_\alpha^2 V}{V} \right), \quad \epsilon\,\equiv\, -\frac{\dot{H}}{H^2}.
\eea
In our setup, these slow roll parameters are small and can be approximated as constant over the time scale of our interest.
Then the propagation eigenstates ($v_1,v_2$) defined as
\begin{eqnarray}
	\left(
	\begin{array}{c}
		a\delta\phi \\
		a\delta\sigma
	\end{array}
	\right)
	=
	\left(
	\begin{array}{cc}
		\cos \theta & \sin\theta \\
		-\sin\theta & \cos\theta 
	\end{array}
	\right)
	\left(
	\begin{array}{c}
		v_1 \\
		v_2 
	\end{array}
	\right),
\end{eqnarray}
obey
\begin{eqnarray}
	\left[ \partial_\tau^2 + (k^2 -\frac{2}{\tau^2} )
	+ 
	\frac{1}{\tau^2} \left( 
	\begin{array}{cc}
	\Delta_{+} & 0 \\
	0 & \Delta_{-}
	\end{array} \right) \right] 
	\left( 
	\begin{array}{c}
		v_1\\
		v_2
	\end{array}
	\right)
	=
	a(\tau) \left( 
	\begin{array}{c}
		S_1\cos\theta -(S_2+S_3)\sin\theta  \\
		 ( S_2 +  S_3)\cos\theta + S_1\sin\theta 
	\end{array}
	\right),
\end{eqnarray}
where the rotation angle $\theta$ is determined as
\begin{eqnarray}
	\sin 2\theta =  -
	\frac{2 \Delta }{\Delta_+ - \Delta_-},
	\quad
	\cos 2\theta =  
	\frac{\Delta_{\phi}-\Delta_{\sigma}}{\Delta_+ - \Delta_-},
\end{eqnarray}
with
\begin{eqnarray}
	\Delta_{\pm}  = \frac{1}{2} (\Delta_\phi + \Delta_\sigma )
	\pm \frac{1}{2} \sqrt{(\Delta_\phi - \Delta_\sigma )^2 + 4 \Delta^2}.
\end{eqnarray}
One can now split the propagation eigenstates into two pieces:
\begin{eqnarray}
	v_1 = v_1^{(v)} + v_1^{(s)},
	\quad
	v_2 = v_2^{(v)} + v_2^{(s)},
\end{eqnarray}
where $v^{(v)}_i$ ($i=1,2$) denote the piece from vacuum fluctuation, while $v^{(s)}_i$ are the piece sourced by gauge fields. Here we are interested in the sourced part which is given by
\begin{eqnarray}
	v_1^{(s)}(\tau,\bfk) &=& 
	\cos\theta \int^\tau d\tau' a(\tau') G_k(\tau,\tau'; \Delta_+) S_1
	-\sin\theta \int^\tau d\tau' \; a(\tau') G_k(\tau,\tau';\Delta_+) (S_2 +S_3),
	\nonumber \\
	v_2^{(s)}(\tau,\bfk) &=&  
	\cos\theta \int^\tau d\tau' \; a(\tau') G_k(\tau,\tau';\Delta_-)( S_2 +S_3)
	+\sin\theta \int^\tau d\tau' \; a(\tau') G_k(\tau,\tau';\Delta_-) S_1,
	\nonumber
\end{eqnarray}
where the Green function $G_k$ obeys 
\begin{eqnarray}
	\left[ 
	\partial_\tau^2 + 
	\left( k^2 - \frac{2 - \Delta_\pm}{\tau^2} \right) \right]
	G_k (\tau,\tau'; \Delta_{\pm}) = \delta(\tau-\tau').
\end{eqnarray}
See Appendix.~\ref{Green} for the properties of this Green function up to first order in slow-roll parameters.

After the dilaton field is stabilized, but before the inflation is over, the dilaton fluctuation $\delta\sigma$ and the source terms $S_i$ are rapidly diluted away, while leaving the inflaton perturbation frozen to be constant in the superhorizon regime. The inflaton perturbation sourced by gauge fields  is determined to be
\begin{eqnarray}
	a(\tau)\delta\phi^{(s)}(\tau,\bfk) \,=\,
	  v_1^{(s)}\cos\theta + v_2^{(s)} \sin\theta\,=\, a \delta\phi^{(S_1)} + a \delta\phi^{(S_2)} + a\delta\phi^{(S_3)},\end{eqnarray}
where
\begin{eqnarray}
	a(\tau)\delta\phi^{(S_1)}&\simeq& \int^\tau d\tau' a(\tau')G_k(\tau,\tau';0) S_1(\tau',\bfk),
	\nonumber \\
a(\tau)\delta\phi^{(S_2)}	&\simeq& \frac{\sin 2 \theta}{2} 
	\int^\tau d\tau' \; a(\tau') 
	[ G_k(\tau,\tau';\Delta_-) - G_k(\tau,\tau';\Delta_+) ] S_2(\tau',\bfk),
	\nonumber \\
a(\tau)\delta\phi^{(S_3)}	&\simeq& \frac{\sin 2 \theta}{2} 
	\int^\tau d\tau' \; a(\tau') 
	[ G_k(\tau,\tau';\Delta_-) - G_k(\tau,\tau';\Delta_+) ] S_3(\tau',\bfk).
	\label{s}
\end{eqnarray}

\begin{figure}[t]
	\centering
	\includegraphics[scale=1]{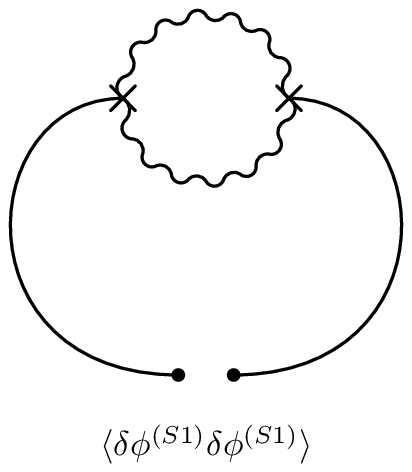}
	\hspace{5mm}
	\includegraphics[scale=1]{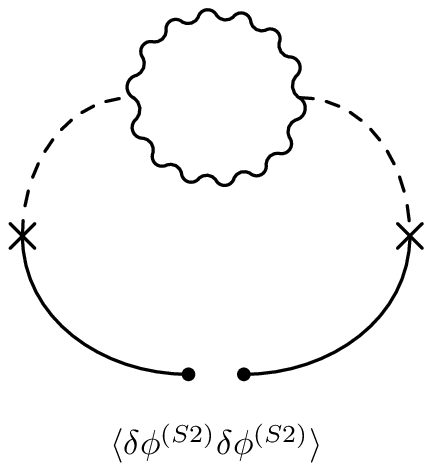}
	\hspace{5mm}
	\includegraphics[scale=1]{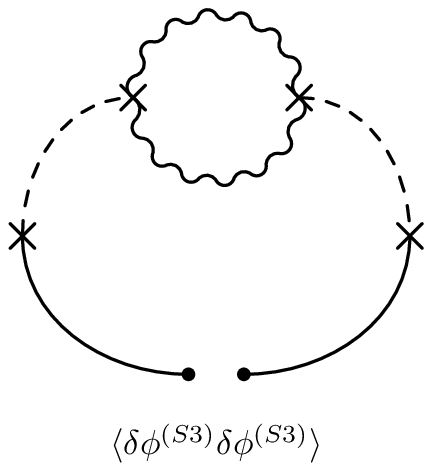}
	\caption{Diagrams of the two-point correlation function of the inflaton perturbation sourced by gauge fields. The solid, dashed, and wavy lines represent the inflaton perturbation $\delta\phi$, the dilaton perturbation $\delta\sigma$, and the gauge field, respectively. The first, second, and third diagrams denote the correlation function sourced by $S_1$, $S_2$, and $S_3$, respectively. The $\times$ mark denotes the gravitational coupling $\alpha_G\sim 1/M_P^2$ accompanying small slow-roll parameter.}
	\label{fig:Diagram}
\end{figure}

In fact, the three source terms are not equally important. We can estimate their relative importance by tracking their dependence on the gravitational coupling $\alpha_G \sim \Mp^{-2}$, as well as investigating the coupling structure of the inflaton and dilaton perturbations. {(See Fig. \ref{fig:Diagram}  for instance.)} It is then straightforward to find  
\bea
\label{S_1}
\frac{\delta \phi^{(S_1)}}{H} &\sim&
\frac{1}{H^3} \left(\frac{\dot\phi_0}{2\Mp^2 H} \right)
 \int d^3\bfp \frac{(k_i-p_i) p_j}{k^2}\widehat{E}_i(\tau,{\bf p}) \widehat{E}_j(\tau,{\bfk - \bf p}) ,
\\
\label{S_2}
\frac{\delta \phi^{(S_2)}}{H} &\sim&
\frac{1}{H^3} \left(\frac{ {\dot\phi}_0  }{2\Mp^2 H} \right) 
\left(\frac{\dot{I}}{HI}\right)
 \int d^3\bfp \widehat{E}_i(\tau,{\bf p}) \widehat{E}_i(\tau,{\bfk - \bf p}) ,
\\
\label{S_3}
\frac{\delta \phi^{(S_3)}}{H} &\sim&
\frac{1}{H^3}\left(\frac{\dot{\phi}_0}{2\Mp^2 H} \right) 
\left(\frac{\dot{\sigma}^2_0}{\Mp^2 H^2} \right)
 \int d^3\bfp \frac{(k_i-p_i) p_j}{k^2}\widehat{E}_i(\tau,{\bf p}) \widehat{E}_j(\tau,{\bfk - \bf p}).
\eea
Note that we are considering the inflaton perturbation after the dilaton
is stabilized. However $\dot\sigma_0$ and $\dot I/I$ in the coefficients
of the above estimates  correspond to the values while the dilaton is rolling, which were approximated as nonzero constants.

Obviously $\delta\phi^{(S_3)}$ is subdominant compared to $\delta\phi^{(S_1)}$ as it is further suppressed by the slow roll parameter $|\dot\sigma_0|/M_PH \ll 1$. On the other hand, as we will see, we need  $n=|\dot I/(HI)|>2$ to enhance the tensor-to-scalar ratio through a rolling dilaton, so the factor $\dot{I}/(HI)$ in (\ref{S_2}) does not cause an additional suppression of $\delta\phi^{(S_2)}$. In fact, from the asymtotic behavior (\ref{EMode}) of the electric mode function, one can easily recognize that the momentum integral of (\ref{S_1})--(\ref{S_2}) for  $|k\tau|\ll 1$ receives the main contribution from the region near $|\bfp|=0$ or $|\bfk-\bfp|= 0$. We then have  schematically 
\bea
\frac{\delta \phi^{(S_1)}(\tau,\bfk)}{\delta \phi^{(S_2)}(\tau,\bfk)} \,\sim\, 
\frac{\int d{\bf q}  \;
\left[{\bf q}\,|\hat{\bfk}-{\bf q}|\right]^{(3/2-n)}}{
\int d{\bf q} \;
\left[{\bf q}\,|\hat{\bfk}-{\bf q}|\right]^{(1/2-n)}},
\eea
where ${\bf q}={\bf p}/k$ is the dimensionless normalized  wave vector. This implies that  the inflaton perturbation sourced by gauge fields is dominated by  $\delta\phi^{(S_2)}$ for the case with $n>2$, where the rolling dilaton can enhance the tensor-to-scalar ratio  significantly. We will therefore consider only $\delta\phi^{(S_2)}$ in the following discussion of scalar perturbation sourced by gauge fields.

\subsection{Scalar power spectrum}
The power spectrum of the inflaton perturbation is defined as
\dis{
\VEV{\delta\phi(\bfk) \, \delta\phi(\bfk ') }
= \frac{2\pi^2}{k^3}\calP_{\delta\phi} (k) \, \delta^{(3)} (\bfk + \bfk ').
\label{Power_phi}
}
In our case, the inflaton perturbation consists of the contribution from vacuum fluctuation  and the piece sourced by gauge fields during the phase of rolling dilaton. Since the sourced part is dominated by $\delta\phi^{(S_2)}$, we have
\dis{
\delta\phi \,\simeq\, \delta\phi^{(v)} + \delta\phi^{(S_2)}.
}
As $\delta\phi^{(v)}$ and $\delta\phi^{(S_2)}$ are uncorrelated, the power spectrum of the curvature perturbation after the dilaton is stabilized is given by
\dis{
\calP_{ {\cal R} } (k) =\calP_{ {\cal R} }^{(v)} (k)+\calP_{ {\cal R} }^{(s)} (k),
}
where $\calP_{ {\cal R} }^{(v)} (k)$ is the nearly scale invariant power spectrum originating from the vacuum fluctuation of the inflaton field:
\bea
	\calP_{ {\cal R}}^{(v)} (k)\,= \,\left(\frac{H}{\dot\phi}\right)^2
	\left(\frac{H}{2\pi}\right)^2 \,\simeq\, \frac{H^2}{8\pi^2\epsilon_\phi \Mp^2},
	\label{PRepsilon}
\eea
and $\calP_{ {\cal R} }^{(s)} (k)$ is the sourced power spectrum:
\dis{
\calP_{ {\cal R} }^{(s)} (k) \,\simeq\, \bfrac{H}{\dot{\phi}}^2 \calP_{\delta\phi}^{(S_2)}(k).
}

Let us now evaluate the sourced power spectrum. Using the solution of $\delta \phi^{(S_2)}$ in (\ref{s}), we find
\bea
\VEV{ \delta\phi^{(S_2)} (\tau, \bfk) \, \delta\phi^{(S_2)} (\tau, \bfk ')}
\,=\, 
\frac{\sin^2 2 \theta}{4 a^2} 
\int^\tau d\tau_1 \, a(\tau_1) [ G_k(\tau,\tau_1;\Delta_-) - G_k(\tau,\tau_1;\Delta_+) ]
\nonumber\\
 \times
\int^\tau d\tau_2 \, a(\tau_2) [ G_{k'}(\tau,\tau_2;\Delta_-) - G_{k'}(\tau,\tau_2;\Delta_+) ]
\VEV{ S_2(\tau_1,\bfk) S_2(\tau_2, \bfk') }.
\label{S2Formula}
\eea
Ignoring the subdominant magnetic field, we find also 
\dis{
&\VEV{S_2(\tau_1,\bfk) S_2(\tau_2,\bfk')}
\simeq 2 a(\tau_1)^2 a(\tau_2)^2 \left(\frac{I_{,\sigma}}{I}\right)^2
\delta^{(3)} (\bfk + \bfk')  \\
&\qquad\times
\int \frac{d^3\bfp}{(2\pi)^3} \left[ 1 + \left( \hat\bfp \cdot\widehat{\bfk-\bfp}\right)^2\right]
{\cal E}(\tau_1,p) {\cal E}(\tau_1,|\bfk-\bfp|) 
{\cal E}^*(\tau_2,p) {\cal E}^*(\tau_2,|\bfk-\bfp|),
}
where ${\cal E} (\tau, k)$ is the electric mode function given in (\ref{EMode}). Then the power spectrum of the sourced curvature perturbation is obtained to be
\bea
{\cal P}_{ {\cal R}}^{(s)} (k)
&\simeq& \frac{2^{4n-2} n^2}{9 \pi^4} \Gamma^4(n+1/2)  \left(\frac{H}{\Mp}\right)^4 
\left\{\int \frac{d^3{\bf q}}{(2\pi)^3} 
\left[ 1 + \left(\hat{\bf q}\cdot\hat{\bf q}'\right)^2\right] q^{-2n+1} q'^{-2n+1} \right.
\nonumber \\
&\times&\left.
\left|
\int^{z_{\ell}}_z dz' z'^{-2n+3}
\left[ 
\left(\frac{2}{3}- \ln \frac{z'}{z} \right) 
+\frac{z^3}{z'^3} \left( \ln \frac{z}{z'} - \frac{2}{3}\right) 
\right] 
\right|^2\right\},
\label{P_int}
\eea
where ${\bf q}\equiv {\bfp}/k$ and ${\bf q}' \equiv (\bfk-\bfp)/k$ are the normalized wave vectors, and $z= -k\tau$. Here the integration over $z'$ is performed from $z$ to $z_{\ell}$, where
\dis{
z_{\ell} =  \frac{1}{{\rm max}(q, q')}
}
corresponds to the time when both 
${\cal E}(\tau,\bfp)$ and ${\cal E}(\tau,\bfk-\bfp) $ 
become a superhorizon mode. 

Regarding to the integration over ${\bf q}$, we note that the electric field stays constant ($n=2$), or grows ($n>2$) in the superhorizon limit $|k\tau|\rightarrow 0$. As a result, the integration suffers from an infrared divergence when the internal momentum approaches to the poles at $qq'=0$.  On the other hand,  only the scales that exit the horizon after the beginning of inflation are relevant for us. Then a physical infrared cutoff at $q_{\rm in}= p_{\rm in} /k$ can be applied to regulate the integral over ${\bf q}$, where $p_{\rm in}$ corresponds to the scale that leaves the horizon at the beginning of  inflation~\cite{Barnaby:2012tk}. Around the region where $q\simeq q_{\rm in}$ or $q'\simeq q_{\rm in}$,  $z_{\ell}$ can be set as $z_{\ell}= 1/q' \simeq 1$ or $z_{\ell}= 1/q \simeq 1$. In the following, we will focus on the case with
\bea
n-2 \,>\, {\cal O}(0.1),
\eea
in which the dilaton-induced gauge fields can significantly affect the primordial perturbations. We then find 
\bea
{\cal P}_{ {\cal R}}^{(s)} (k)
\simeq \frac{2^{4n} n^2}{27 \pi^6} \Gamma^4(n+1/2)  \left(\frac{H}{\Mp}\right)^4 
\frac{q_{\rm in}^{-2n+4}}{2n-4}
\left|
\frac{9 z^{-2n+4}}{4 ( 2n-1)^2 (n-2)^2}
\right|^2.
 \label{PRs}
\eea
Note that this power spectrum has an explicit scale dependence.  
This is because 
 the inflaton perturbation gets affected by the growing electric modes even after it exits the horizon.

If this mechanism continues until the end of inflation, the dilaton and gauge field fluctuations eventually lead to a too large deviation of the scalar spectral index from the observed value, and/or a too large non-gaussainity in the curvature perturbation. To avoid it, we assume that the dilaton rolling  is terminated at some time $\tau_D$ due to the stabilization which  is accomplished within a short time interval  $\Delta t \ll H^{-1}$. We then apply the instantaneous stabilization approximation in which the inflaton perturbation $\delta\phi$  at $\tau_D$ is matched to the frozen solution in the absence of source terms. In this approach, the normalized conformal time $z=-k\tau$ in (\ref{PRs}) can be replaced by
\dis{
z_D\,\equiv\, -k\tau_D \,=\, e^{-N_k + N_D},\label{zD}
}
where $N_k$  denote the number of e-foldings from  the horizon exit ($\tau_k=-1/k$) to the end of inflation, while $N_D$ is the number of e-foldings from the dilaton stabilization. By definition, $p_{\rm in}$ corresponds to the scale that leaves the horizon at the beginning of inflation, so
\dis{
q_{\rm in} \,=\, p_{\rm in}/k \,=\,  e^{-N_T + N_k}\label{qin},
}
where $N_T$ is the total number of e-foldings over the inflation epoch. {See figure.~\ref{fig:efolding} for the e-folding numbers relevant for our setup}. Putting (\ref{zD}) and (\ref{qin}) together, the sourced power spectrum (\ref{PRs}) can be expressed in terms of the e-folding numbers as
\dis{
{\cal P}_{ {\cal R}}^{(s)} (k)
\,\simeq&\,\, \frac{2^{4n} n^2}{27 \pi^6} \Gamma^4(n+1/2)  \left(\frac{H}{\Mp}\right)^4 
\frac{e^{(2n-4)(N_T-N_k)}}{(2n-4)}
 \left|
\frac{9 e^{(2n-4) (N_k - N_D)} }{4(2n-1)^2(n-2)^2}
\right|^2 
 \label{PRsN}
}
for $(n-2) > {\cal O}(0.1)$.

\begin{figure}[t]
	\centering
	\includegraphics[scale=1]{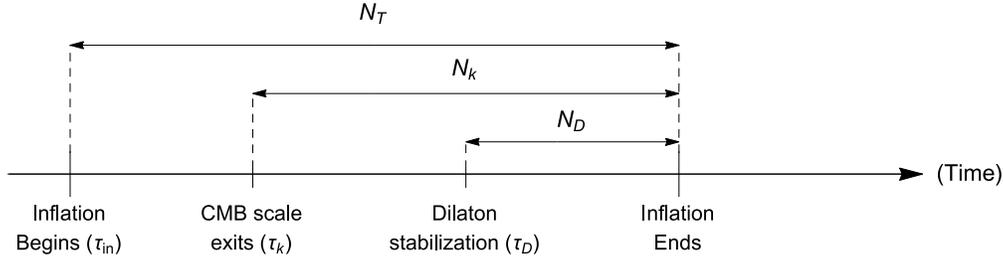}
	\caption{{Relevant e-folding numbers in our scenario involving  the beginning of inflation which is assumed to take place at a similar time as the beginning of dilaton rolling, horizon exit of the CMB scales,  stabilization of the dilaton, and the end of inflation.}}
\label{fig:efolding}
\end{figure}

As the sourced power spectrum has an explicit scale dependence, it should be tightly constrained by observations which indicate that the curvature perturbation is nearly scale invariant.   Using the above results, we find the spectral index for the total curvature power spectrum\footnote{There is also a model dependent contribution to $n_s$ from the $k$-dependence of the dilaton coupling evolution rate $n$. For the model considered in section \ref{model}, we find  $d\ln n/d\ln k \sim \epsilon$, which is negligible compared to the contribution from the $k$-dependence of $N_k$. } is given by
\bea\label{ns}
n_s-1 \simeq (n_s-1)^{(v)} - 2 \left( \frac{  {\cal P}_{\cal R}^{(s)}/{\cal P}_{\cal R}^{(v)}  }{1+{\cal P}_{\cal R}^{(s)}/{\cal P}_{\cal R}^{(v)}}\right) \Big[ (n-2) + \eta_\phi - \epsilon \Big],
\eea
where  $(n_s-1)^{(v)} = 2\eta_\phi - 6\epsilon$ is the spectral index for the scalar perturbation generated by vacuum fluctuation. According to the recent Planck observation~\cite{Ade:2015lrj},  
$$n_s = 0.9655 \pm 0.0062, $$ 
at 68\% confidence level. Assuming that $n_s^{(v)}$ is not too far away from this observed value, e.g.  $(n_s-n_s^{(v)}) \,\lesssim\, 5\times 10^{-2}$, the sourced scalar power spectrum is constrained as
\dis{
{\cal P}_{\cal R}^{(s)} / {\cal P}_{\cal R}^{(v)} \, \lesssim\,  \frac{5\times 10^{-2}}{2(n-2)}
\label{const_scale}
} 
for $(n-2)> {\cal O}(0.1)$.
In the following, we will use this as an observational constraint on the scalar power spectrum sourced by rolling dilaton.

In addition to explicit scale dependence, our setup can give rise to an anisotropic signal in the scalar power spectrum~\cite{Bartolo:2012sd}. Although we do not assume any pre-existing background gauge field, the dilaton-induced  gauge field modes stretched far beyond the scale of observable universe can be considered as a background field having a preferred direction. Such large scale vector field may lead to an anisotropic signal in the scalar power spectrum as
\dis{
{\cal P}_{\cal R}({\bf k}) = {\cal P}^{(v)}_{\cal R}(k)\big(1 + g_* \cos^2\theta \big),
}
where $g_{*}$ is constrained by the recent Planck data~\cite{Ade:2015lrj}, 
$$	g_{*} \simeq (0.23^{+1.70}_{-1.24})\times 10^{-2}	$$
at the $68\%$ confidence level. See also~\cite{Dulaney:2010sq,Gumrukcuoglu:2010yc,Watanabe:2010fh,Ohashi:2013qba} for the discussion of statistical anisotropy in the presence of initial background gauge field. In our scenario,  $ g_{*} \sim {\cal P}_{\cal R}^{(s)}/{\cal P}_{\cal R}^{(v)} $ up to ${\cal O}(1)$ numerical factor. If we require the amplitude of sourced scalar power spectrum as 
\bea
 {\cal P}_{\cal R}^{(s)}/{\cal P}_{\cal R}^{(v)} \lesssim 0.04,
 \label{const_scale_aniso}
\eea
then this bound covers both the constraint from the observed spectral index and the constraint from the anisotropic signal at $2\sigma$ level.

\subsection{Scalar bispectrum}

We have seen that an additional scalar perturbation is generated by rolling dilaton field.  It has been noticed  that the curvature perturbation due to the fluctuation of unstabilized dilaton  has a non-gaussian distribution with a nearly local shape~\cite{Barnaby:2012tk,Fujita:2013qxa}. Even after the dilaton field is stabilized, the inflaton perturbation which was originated from the source terms $S_i$ may still include a non-gaussian piece.

As the inflaton perturbation $\delta\phi^{(v)}$ from  vacuum fluctuation
is nearly gaussian, the leading scalar bispectrum comes from the sourced perturbation $\delta\phi^{(S_2)}$, i.e.
\bea
&&\VEV{ \delta\phi(\tau,\bfk_1) \delta\phi(\tau,\bfk_2) \delta\phi(\tau,\bfk_3)} \nonumber \\
&\simeq &
\left( \frac{\sin 2\theta}{2a} \right)^3
 \int \prod_{l=1}^3d\tau_l a(\tau_l)\left[G_{k_l}(\tau,\tau_l;\Delta_-) - G_{k_l}(\tau,\tau_l;\Delta_+) \right]
\VEV{S_2(\tau_1,\bfk_1) S_2(\tau_2,\bfk_2) S_2(\tau_3,\bfk_3)}.\nonumber
\eea
Using the gauge mode function (\ref{EMode}) and the Green function (\ref{GDifference}), we find
\bea
&&\VEV{ \delta\phi(\tau,\bfk_1) \delta\phi(\tau,\bfk_2) \delta\phi(\tau,\bfk_3)} \nonumber\\
&&=
\left(-\frac{2^{2n} n \dot{\phi}}{3\pi} \Gamma^2(n+1/2) \frac{H}{\Mp^2}  \right)^3 
\left\{
\int \prod_{l=1}^3\frac{d^3{\bfp}_l}{(2\pi)^{3/2}} p_l^{-2n+1} \right.
\nonumber\\
&&\quad\times
 \int \prod_{l=1}^3d\tau_l\tau_l^{-2n+3} 
\left[
\left(\frac{2}{3} - \ln \frac{\tau_l}{\tau} \right) +
\frac{\tau^3}{\tau_l^3}\left(\ln \frac{\tau}{\tau_l} - \frac{2}{3} \right)
\right]
\nonumber \\
&&\quad\times\ 
\delta^{(3)}(\bfp_1 + \bfk_2 - \bfp_2)
\delta^{(3)}(\bfp_2 + \bfk_3 - \bfp_3)
\delta^{(3)}(\bfp_3 + \bfk_1 - \bfp_1)
\nonumber \\
&&\quad\times\left.\frac{}{}
\left[ ({\hat \bfp}_1 \cdot {\hat \bfp}_2)^2 
+ ({\hat \bfp}_2 \cdot {\hat \bfp}_3)^2 
+ ({\hat \bfp}_3 \cdot {\hat \bfp}_1)^2 - 
({\hat \bfp}_1 \cdot {\hat \bfp}_2)({\hat \bfp}_2 \cdot {\hat \bfp}_3)({\hat \bfp}_3 \cdot {\hat \bfp}_1) \right]
\right\},
\label{3pt}
\eea
where the contribution from the magnetic mode  is ignored as it is subdominant for positive $n$. As in the case of power spectrum, we can introduce the infrared cutoff $p_{\rm in}$ for the momentum integral, being the scale leaving the horizon at the beginning of inflation. Then the 3-point function of the curvature perturbation is obtained to be
\bea
\label{curvature-3point}
	&& \langle {\cal R}_\phi(\tau,\bfk_1) {\cal R}_ \phi(\tau,\bfk_2)
	{\cal R}_\phi(\tau,\bfk_3)
	\rangle \nonumber \\
	 &\simeq & 
	\frac{1}{(2\pi)^{9/2}}
	\left[
	-\frac{3 \cdot 2^{2n-2}  \Gamma^2(n+1/2) }{\pi } 
	\frac{n \, z^{-2n+4}}{ (2n-1)^2 (n-2)^2}
	\right]^3
	\nonumber \\
	&\times&
	\frac{q_{\rm in}^{-2n +4}}{2n-4}
	\left( \frac{H}{\Mp} \right)^6
	\delta^{(3)}(\bfk_1 + \bfk_2 + \bfk_3)
	\nonumber\\
	&\times&
	\frac{8\pi}{3}
	\left[ 
		\frac{1+(\hat{\bfk}_1 \cdot\hat{\bfk}_2)^2}{k_1^3k_2^3} +
		\frac{1+ (\hat{\bfk}_2 \cdot\hat{\bfk}_3)^2}{k_2^3k_3^3} + 
		\frac{1+(\hat{\bfk}_3 \cdot\hat{\bfk}_1)^2}{k_3^3k_1^3} \right],
\eea
which has a nearly local shape~\cite{Barnaby:2012tk}. The corresponding non-linearity parameter $\fnl$ is given by
\bea
	f_{NL} \,\simeq\, -\frac{20}{27} \epsilon^3_\phi {\cal P}^{(v)}_{\cal R}(k)
	\left[ \frac{e^{(2n-4)(N_T-N_k)}}{2n-4}\right]
	\left[
	\frac{3 \cdot 2^{2n}  \Gamma^2(n+1/2) }{\pi } 
	\frac{n \, e^{(2n-4)(N_k-N_D)}}{ (2n-1)^2 (n-2)^2}
	\right]^3
	\label{fnl}
\eea
for $(n-2) > {\cal O}(0.1)$, where we expressed $z=z_D$ and $q_{\rm in}$ in (\ref{curvature-3point}) in terms of the e-folding numbers as in the case of the scalar power spectrum. The recent Planck data provides a strong bound on the non-gaussianity~\cite{Ade:2015lrj}, implying 
\bea
|f_{NL}| \,\lesssim \, 10,
\label{non-gaussianity-constraint}
\eea
which should be applied to the above non-gaussianity due to the scalar perturbation sourced by rolling dilaton.

\section{Tensor perturbation} \label{tensorpert}

\subsection{Tensor power spectrum}
Gravitational wave corresponds to the traceless-transverse component of the metric perturbation, and  obeys the equation of motion 
\begin{eqnarray}
	h_{ij}'' + 2 {\cal H} h_{ij}' + k^2 h_{ij} = \frac{2}{\Mp^2}	T_{ij}^{(TT)},
\end{eqnarray}
where $T_{ij}^{(TT)}$ denotes the traceless-transverse component of the energy momentum tensor. After the Fourier expansion  
\begin{eqnarray}
	h_{ij}(\tau, \bfx) &=&
	\int \frac{d^3 \bfk}{(2\pi)^{3/2}} \sum_{\lambda} \Pi_{ij,\lambda}(\hat \bfk) 
	\hat{h}_\lambda (\tau,\bfk) e^{i \bfk \cdot \bfx},
	\\
	{\hat h}_\lambda(\tau,\bfk) &=&
	h_\lambda(\tau,k) a_{\bfk, \lambda} +
	h^*_\lambda(\tau,k) a^\dagger_{-\bfk,\lambda},
\end{eqnarray}
where $\Pi_{ij,\lambda}(\hat \bfk) = \epsilon_{i,\lambda}(\hat \bfk)\epsilon_{j,\lambda}(\hat \bfk)$ is the traceless-transverse polarization tensor, we find the equation of motion for the normalized tensor mode $Q_\lambda \equiv \frac{\Mp}{2} a  h_\lambda$ is given by
\begin{eqnarray}
	Q_\lambda'' + \left( k^2 - \frac{a''}{a} \right) Q_\lambda
	= S_\lambda(\tau,\bfk),
	\label{TensorEQ}
\end{eqnarray}
where  
\dis{
S_\lambda(\tau,\bfk)= \frac{a}{\Mp} 
	\int \frac{d^3 {\bf x}}{(2\pi)^{3/2}} e^{-i\bfk \cdot \bfx} 
	\, \Pi^*_{ij,\lambda}(\hat \bfk) \, T_{ij}^{(TT)}.
	\label{Slambda}
}
As usual, the solution of \eqref{TensorEQ} is given by the sum of the homogeneous solution (vacuum part) and a particular solution for the source $S_\lambda$. In our setup,  the gauge fields produced by rolling dilaton  contribute to the energy momentum tensor as
\begin{eqnarray}
	T_{ij}^{U(1)} = a^2 [ \rho_{\rm U(1)} \delta_{ij} - (E_iE_j +B_iB_j)],
\end{eqnarray}
where $ \rho_{U(1)}$ is given in \eqref{rhoem}. Plugging this energy momentum tensor into \eqref{Slambda}, we find 
\begin{eqnarray}
	S_\lambda(\tau,\bfk) = 
	-\frac{a^3}{\Mp} \Pi_{ij,\lambda}^* (\hat\bfk) 
	\int \frac{d^3 \bfp}{(2\pi)^{3/2}} 
	\left( 
	\hat{E}_i(\tau,{\bf p})  \hat{E}_j(\tau,{\bf k} - {\bf p})
	+ \hat{B}_i(\tau,{\bf p})  \hat{B}_j(\tau,{\bf k} - {\bf p})
	\right).
	\label{TensorSource}
\end{eqnarray}

Similar to the scalar power spectrum  \eqref{Power_phi}, the tensor power spectrum  ${\cal P}_\lambda(k)$ is defined as
\dis{
\VEV{h_\lambda(\bfk) h_\lambda(\bfk')} = \frac{2\pi^2}{k^3}{\cal P}_\lambda(k) \delta^{(3)}(\bfk+\bfk'),
}
where $h_\lambda(\bfk)$ consists of two pieces, the vacuum fluctuation $h^{(v)}_\lambda(\bfk)$ and the additional  fluctuation $h^{(s)}_\lambda(\bfk)$ sourced by gauge fields. Since these two pieces are uncorrelated to each other, the tensor power spectrum is simply given by
\dis{
{\cal P}_\lambda(k) \,= \, \calP_\lambda^{(v)} (k) + {\cal P}^{(s)}_{\lambda}(k),
}
where
${\cal P}^{(v)}_{\lambda}(k) =  {H^2}/{\pi^2 \Mp^2}$
is the usual tensor spectrum from the vacuum fluctuation. The sourced part  can be obtained by solving  \eqref{TensorEQ}, which yields
\bea
&&	\delta^{(3)}(\bfk + \bfk') {\cal P}^{(s)}_{\lambda}(k) \nonumber \\
	&=&  
	\left( \frac{2}{a\Mp} \right)^2
	\left( \frac{k^3}{ 2\pi^2} \right) 
	\int d\tau_1 G_k(\tau,\tau_1) \int d\tau_2 G_{k'}(\tau,\tau_2)
	\langle S_\lambda (\tau_1,{\bf k}) S_\lambda (\tau_2, {\bf k'}) \rangle, \nonumber
\eea
where 
\dis{
G_k(\tau,\tau') \equiv G_k(\tau,\tau' ; 0) \simeq \frac{1}{k^3 \tau\tau'} \Big[ k\tau' \cos k\tau' - \sin k\tau' \Big] \quad {\rm for} \quad -k\tau \ll1. \nonumber
} 

Repeating the same procedure as we did to get the scalar power spectrum, we find the following general formula for the sourced tensor power spectrum: 
\bea
{\cal P}^{(s)}_\lambda (k) \simeq
\frac{k^3}{a^2 \pi^2} \frac{1}{\Mp^4}
\int \frac{d^3\bfp}{(2\pi)^3} 
\left[1+ (\hat{\bfk} \cdot \hat{\bfp})^2 \right] \left[ 1 + (\hat{\bfk} \cdot \widehat{\bfk-\bfp})^2 \right]
\nonumber \\
\times
\left| 
\int^\tau d\tau' a^3(\tau') G_k(\tau,\tau')  {\cal E}(\tau',p) {\cal E}(\tau',|\bfk-\bfp|) 
\right|^2,
\eea
where the subdominant contribution from the magnetic modes is ignored. Using the solution  \eqref{EMode}, and also introducing the infrared cutoff $p_{\rm in}$ (see \eqref{qin}) for the momentum integration, the sourced tensor power spectrum is obtained to be
\bea
{\cal P}_t^{(s)}&=& \sum_\lambda {\cal P}_\lambda^{(s)}\nonumber \\
	&\simeq& 
	\frac{2^{4n+1}\Gamma^4(n+1/2)}{27\pi^6} \left( \frac{H}{\Mp} \right)^4
	\left[ \frac{e^{(2n-4)(N_T-N_k)}}{2n-4} \right]
	\left[\frac{e^{(2n-4)(N_k-N_D)}  }{2n-4}  \right]^2
	 \label{PRt}
\eea
for $(n-2)\gg {\cal O}(\epsilon)$. Then the spectral index of ${\cal P}_t \,=\, {\cal P}_t^{(v)}+{\cal P}_t^{(s)}$ is given by 
\bea
n_t \,\simeq\,  n_t^{(v)} - 2 \left(\frac{ {\cal P}_t^{(s)}/{\cal P}_t^{(v)}}{1+ {\cal P}_t^{(s)}/{\cal P}_t^{(v)}} \right) 
\Big[(n-2)+\epsilon\Big],
\eea
where $ n_t^{(v)} = -2  \epsilon$ is the spectral index for the tensor power spectrum of the vacuum fluctuation. If the sourced tensor power spectrum dominates over the vacuum contribution, i.e. ${\cal P}_t^{(s)}/{\cal P}_t^{(v)}\gtrsim 1$,  which is the case of our prime interest, the tensor spectral index is approximately given by
\dis{
n_t \,\simeq\, -2(n-2),
}
and therefore the tensor spectrum in our setup can be a lot more red-tilted compared to the standard result $n_t^{(v)}=-2\epsilon$.

There also exists non-vanishing cross correlation between curvature and tensor perturbation induced by the dilaton-induced gauge fields. In our setup, such correlation is estimated as
\bea
\langle h {\cal R} \rangle \sim \sqrt{r g_*} \langle {\cal R R} \rangle
\eea
so is bounded by the anisotropy factor, $g_* \lesssim {\cal O}(10^{-2})$, as well as by tensor-to-scalar ratio, $r \lesssim 0.1$. This cross correlation does not affect the temperature power spectrum \cite{Contaldi:2014zua,Zibin:2014iea,Emami:2014xga}. Instead, with the anisotropic part of scalar power spectrum, they contribute to quadrupole anisotropy in the temperature, which is the correlation between $\ell$ and $\ell + 2$. At least for the case $g_* \sim {\cal O}(0.01)$ and $r \lesssim {\cal O}(0.01)$, the contribution from scalar-tensor correlation would be smaller than the contribution from anisotropic part of curvature power spectrum. In this case, the observational constraint from quadrupole anisotropy is already taken into account through $g_*$. Despite of this, it is interesting to see how the scalar-tensor correlation affects quadrupole anisotropy in detail, and how the constraint from statistical anisotropy changes when the scalar-tensor correlation becomes important. We leave this for future work.

\subsection{Tensor-to-scalar ratio}
In the standard single field inflation scenario, the tensor-to-scalar ratio is determined by the inflaton slow-roll parameter $\epsilon_\phi$ as
\dis{
r^{(v)} = \frac{ {\cal P}_t^{(v)}} {{\cal P}_{\cal R}^{(v)}} = 16 \epsilon_\phi.
}
On the other hand, in our setup the gauge fields produced by rolling \dilaton contribute to both the scalar and tensor power spectra. Our results  \eqref{PRsN} and \eqref{PRt}, including the contributions from rolling dilaton, are summarized as
\begin{eqnarray}
	{\cal P}_{\cal R}(k) &\simeq& 
	{\cal P}^{(v)}_{\cal R} (k) 
	\left[ 1 + \epsilon^2_\phi \, {\cal P}_{\cal R}^{(v)}(k) f_{\cal R}(n,k) \right],
	\nonumber \\
	{\cal P}_{t}(k) &\simeq& 
	16\epsilon_\phi \, {\cal P}^{(v)}_{\cal R} (k) 
	\left[1 + \epsilon_\phi \, {\cal P}^{(v)}_{\cal R} (k) f_t(n,k) \right],
\end{eqnarray}
where 
\bea
{\cal P}_{\cal R}^{(v)} (k) \,=\, \frac{H^2}{8\pi^2 \epsilon_\phi M_P^2},\eea
and 
\bea
f_{\cal R} (n,k)&=&  \Gamma^4(n+1/2)  \frac{2^{4n+6} n^2}{27\pi^2}
\left[ \frac{e^{(2n-4)(N_T-N_k)}}{2n-4}  \right]
\left[ \frac{9 e^{(2n-4)(N_k-N_D)}}{4(2n-1)^2(n-2)^2} \right]^2,
\\
f_t (n,k) &=& \Gamma^4(n+1/2)  \frac{2^{4n+4}}{27\pi^2} \left[ \frac{e^{(2n-4)(N_T-N_k)}}{2n-4} \right]
\left[ \frac{e^{(2n-4)(N_k-N_D)}}{2n-4} \right]^2,
\eea
for $(n-2) > {\cal O}(0.1)$. The resulting tensor-to-scalar ratio is given by\footnote{The tensor-to-scalar ratio was calculated also in \cite{Ohashi:2013qba} when the dilaton field couples to the gauge field. Ref. \cite{Ohashi:2013qba} focused on observational signatures of models when $n=2$ and the dilaton field continues to roll until the end of inflation. In this case, the energy density of gauge field stays nearly constant. On the other hand, we are discussing a different situation with $n>2$, where the energy density of gauge field continuously grows in the superhorizon limit until when the dilaton is stabilized, which is assumed to take place before the end of inflation. This is why our result on tensor-to-scalar ratio is different from \cite{Ohashi:2013qba}.  }
\bea
r \,=\, \frac{ {\cal P}_{t} }{ {\cal P}_{\cal R} } 
\,=\, 16\epsilon_\phi 
\left(
\frac{1 + \epsilon_\phi \,f_t(n,k) \; {\cal P}_{\cal R}^{(v)} }
{1 + \epsilon^2_\phi \, f_{\cal R}(n,k) \; {\cal P}_{\cal R}^{(v)} }
\right). \label{tensor-r}
\eea

\begin{figure}[t]
	\centering
	\includegraphics[scale=.8]{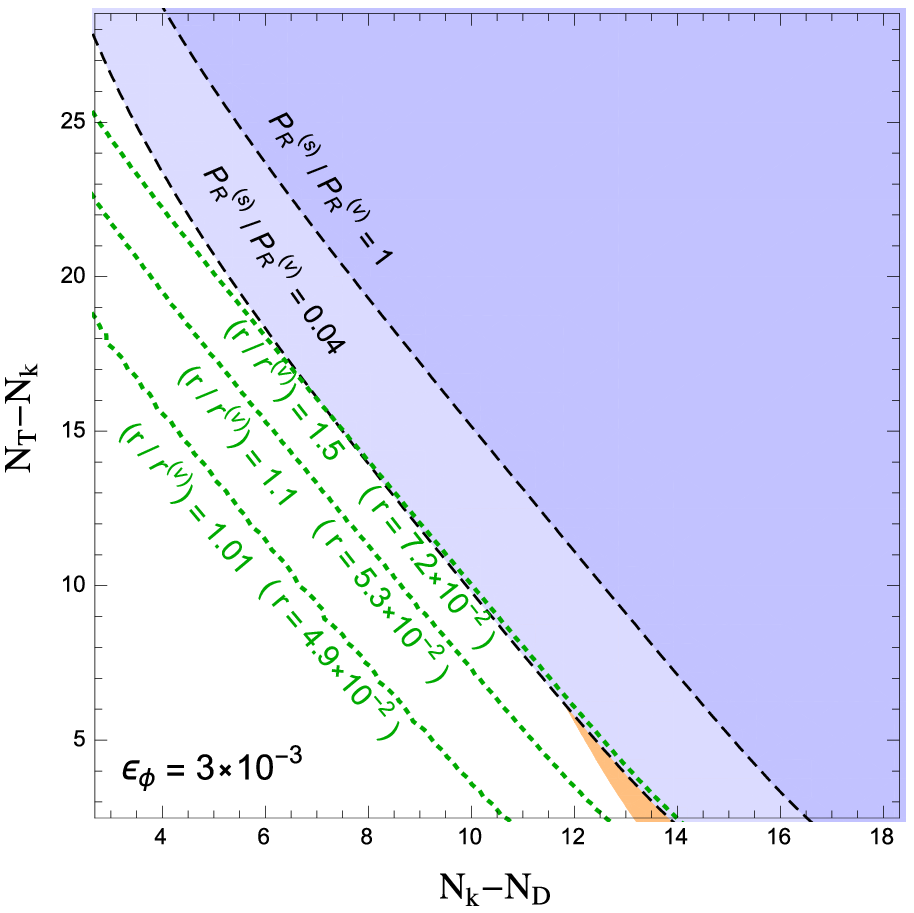}
	\hspace{.5cm}
	\includegraphics[scale=.8]{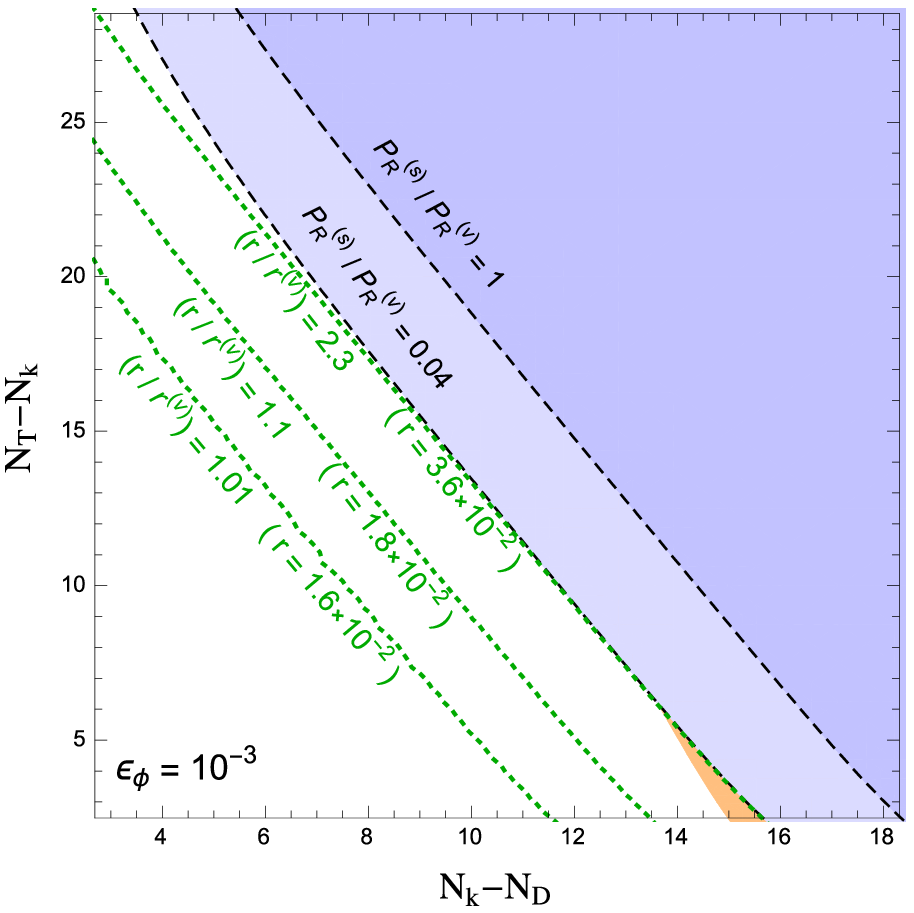}
	\includegraphics[scale=.8]{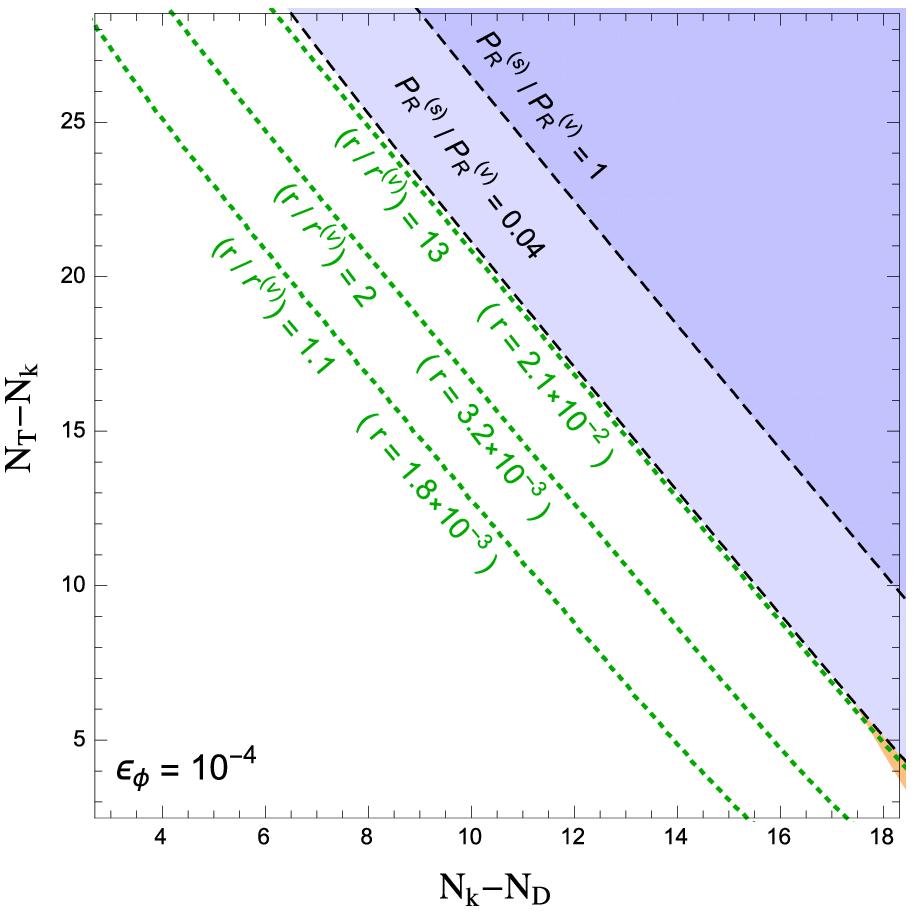}
	\hspace{.5cm}
	\includegraphics[scale=.8]{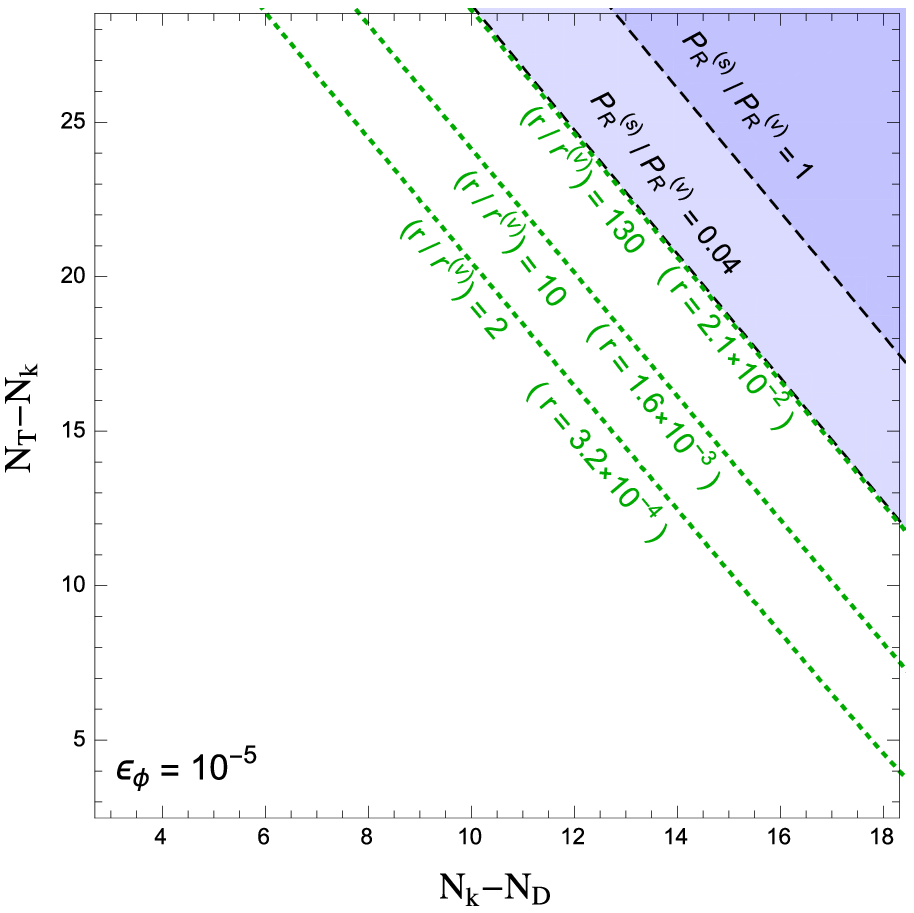}
	\caption{ {The contour plots show $r$ (and also the corresponding  $r/r^{(v)}$) for $n=2.3$ and $\epsilon_\phi= (3\times 10^{-3},\, 10^{-3},\, 10^{-4},\,10^{-5})$. The blue-shaded region on the upper right corner is disfavoured by giving a too large deviation of the scalar spectral index from the observed value, while  the orange-shaded region gives $|f_{NL}| \gtrsim 10$  disfavoured by the recent Planck data.}}
\label{fig:r}
\end{figure}

In figure~\ref{fig:r}, we show the allowed value of the tensor-to-scalar ratio $r$, as well as the corresponding  enhancement factor $r/r^{(v)}$, on the plane of $(N_T-N_k, N_k-N_D)$ for $n=2.3$ and the four different values of the inflaton slow roll parameter $\epsilon_\phi = ( 3\times 10^{-3}, \,10^{-3},\, 10^{-4},\, 10^{-5})$. The numerical results depicted in figure~\ref{fig:r} and other figures in the following  include the subleading corrections which were ignored in the approximate analytic results in (\ref{PRsN}) and (\ref{fnl}). We recall that $n\equiv -\dot{I}/HI$ is the evolution rate of the dilaton-dependent gauge coupling $g(\sigma)=1/I(\sigma)$, $\epsilon_\phi$ is related to the inflationary Hubble scale as 
\bea
\epsilon_\phi \,\simeq\, 0.01 \left(\frac{H}{10^{14}\, {\rm GeV}}\right)^2,
\eea 
and $N_T-N_k$ corresponds to the number of e-foldings from the beginning of inflation to the horizon exit, while $N_k-N_D$ is the number of e-foldings from the horizon exit to the dilaton stabilization. {(See figure.~\ref{fig:efolding}.)

 The blue-shaded part in figure~\ref{fig:r} corresponds to the region with ${\cal P}_{\cal R}^{(s)}/{\cal P}_{\cal R}^{(v)} \,\gtrsim\, 0.04$, which is in conflict with the constraint (\ref{const_scale}) from the observed scalar spectral index. On the other hand, the orange-shaded region is disfavoured as it gives $|f_{NL}| \gtrsim 10$ which is in conflict with the recent Planck data~\cite{Ade:2015ava}. Our results show that $r/r^{(v)}={\cal O}(1)$  for $\epsilon_\phi ={\cal O}(10^{-3})$. On the other hand, for smaller $\epsilon_\phi$, observational constraints to prevent a large enhancement of $r$ are weakened, which can be noticed from the $\epsilon_\phi$-dependence of the sourced scalar power spectrum in \eqref{tensor-r}. As a consequence, for $\epsilon_\phi\ll 10^{-3}$, $r/r^{(v)}\gg 1$ can be achieved without any conflict with the observational constraints. This makes it possible that $r\sim10^{-2}$ even when $\epsilon_\phi\ll 10^{-3}$.}

\begin{figure}[t]
	\centering
	\includegraphics[scale=.8]{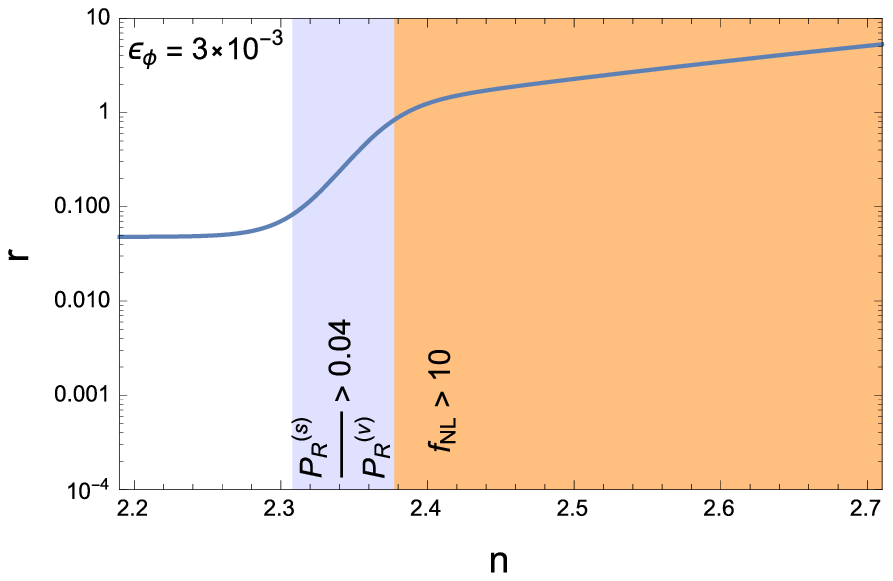}
	\hspace{.5cm}
	\includegraphics[scale=.8]{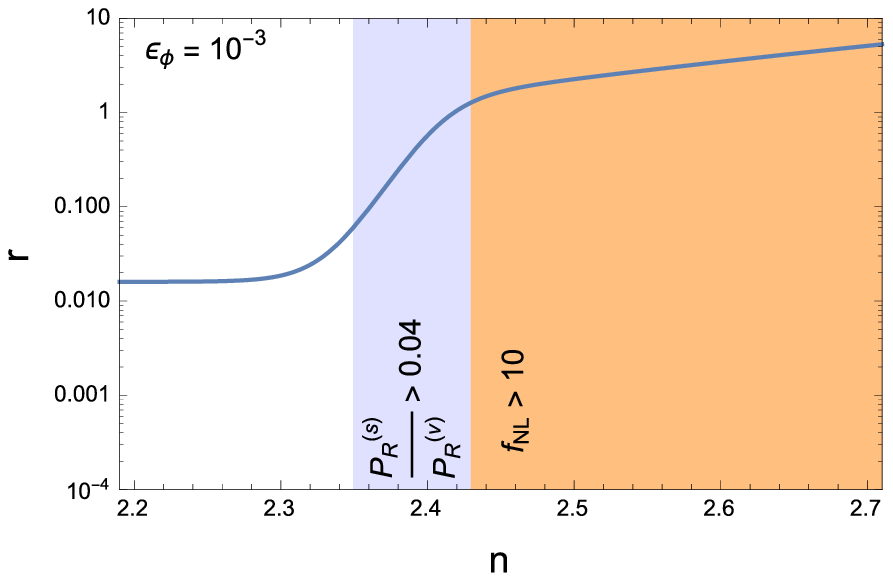}
	\includegraphics[scale=.8]{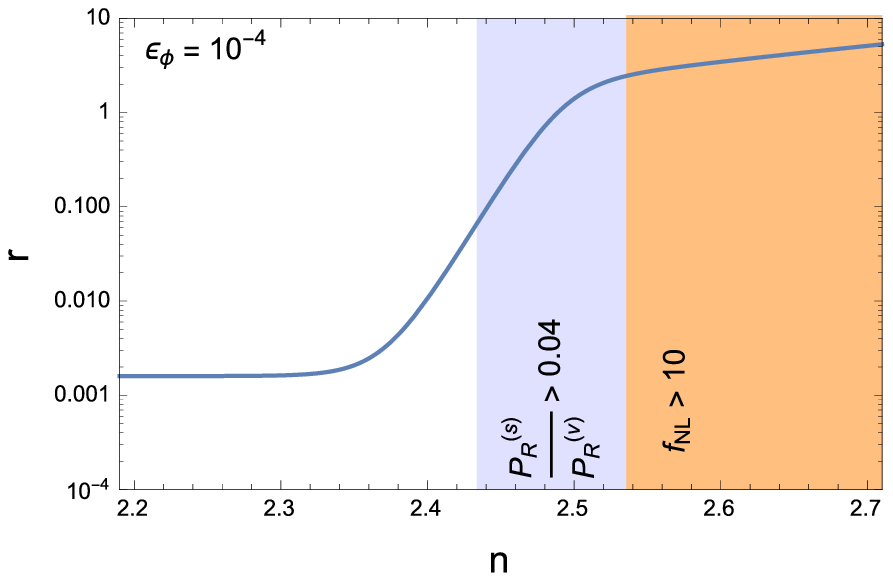}
	\hspace{.5cm}
	\includegraphics[scale=.8]{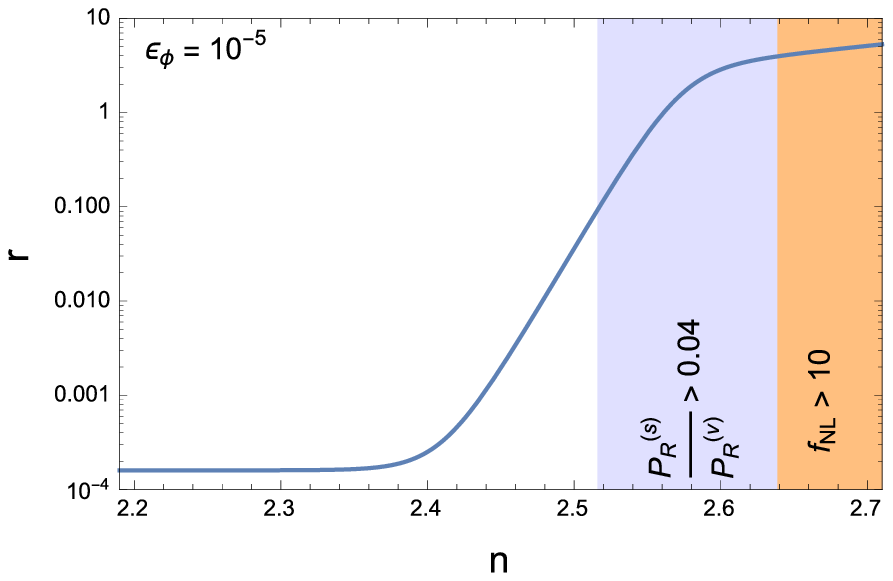}
	\caption{The values of $r$ as a function of $n$ for $(N_k-N_D,N_T-N_k )=(5,20)$ and $\epsilon_\phi = (3\times 10^{-3},\, 10^{-3},\, 10^{-4},\, 10^{-5})$. Again the orange region is disfavoured  by giving a too large non-gaussianity $(\fnl \gtrsim 10)$, and the blue region is disfavored by giving a too large deviation of the scalar spectral index from the observed value.
}
\label{fig:r_n}
\end{figure}

In figure~\ref{fig:r_n},  we plot the tensor-to-scalar ratio $r$ as a function of $n$ for $(N_k-N_D,\, N_T-N_k )=(5,20)$ and $\epsilon_\phi =(3 \times 10^{-3},\, 10^{-3},\, 10^{-4},\,10^{-5})$. Again the blue-shaded part denotes the region with ${\cal P}_{\cal R}^{(s)}/{\cal P}_{\cal R}^{(v)} \gtrsim 0.04$, and  the orange-shaded region gives  $|f_{NL}| \gtrsim 10$.

From our results, we can notice that the allowed maximal value of $r$ is rather insensitive to the value of $\epsilon_\phi$. This can be understood  from that the spectral index constraint \eqref{const_scale_aniso} leads to
\bea
r &=&  \frac{{\cal P}_t^{(v)}+{\cal P}_t^{(s)}}{{\cal P}_{\cal R}^{(v)}+{\cal P}_{\cal R}^{(s)}} \,=\, \frac{16\left[\epsilon_\phi + (f_t/f_{\cal R})({\cal P}_{\cal R}^{(s)}/{\cal P}_{\cal R}^{(v)})\right]}{1+{\cal P}_{\cal R}^{(s)}/{\cal P}_{\cal R}^{(v)}}
\nonumber \\
&\lesssim & 16 \frac{f_t}{ f_{\cal R} } \frac{0.04}{1+0.04} 
\,\simeq \, 0.64\,\frac{(2n-1)^4(n-2)^2}{(9n)^2}
\label{UB}
\eea
and this $\epsilon_\phi$-independent upper bound on $r$ can be nearly saturated in most cases.

\section{A model}
 \label{model}

In this section, we discuss a model which can realize the scenario discussed in the previous sections. As a specific example, we consider the dilaton potential and gauge kinetic function given by
\bea
V(\sigma) \,\simeq \, \mu^3\sigma + \mbox{dilaton-independent part}, \quad
I(\sigma) \,=\, e^{\sigma/\Lambda},
\label{dilaton-potential} \eea
where $\mu$ and $\Lambda$ are constant  mass parameters. The above linear dilaton potential is valid only for $\sigma$ in the rolling regime, and we assume that the dilaton field is quickly stabilized by a steep potential (see fig. \ref{fig:Dilaton potential}) after the rolling period is over.

During the rolling period, the background dilaton field $\sigma_0$ obeys
\begin{eqnarray}
	\ddot{\sigma}_0 + 3 H \dot{\sigma}_0 + \mu^3 \,\simeq\, 0,
\end{eqnarray}
where we have neglected the back-reaction effect from the dilaton coupling to gauge fields. Assuming a slow-roll motion, the solution  is given by
\begin{eqnarray}
	\sigma_0(t) = \sigma_{\rm in}-\frac{\mu^3}{3H} (t-t_{\rm in}),
	\label{Apsi}
\end{eqnarray}
where $\sigma_{\rm in}$ and $t_{\rm in}$ denote the field value and the Robertson-Walker time coordinate, respectively,  at the beginning of the rolling. Substituting \eqref{Apsi} to the gauge kinetic function $I(\sigma_0)$ in \eqref{dilaton-potential}, we find
\begin{eqnarray}
	I(t) \,=\,e^{\sigma_{\rm in} / \Lambda } \;  (a/a_{\rm in})^{-n}, 
	\quad n \,\equiv \, -{\dot{I}}/{HI}\,=\, {\mu^3}/{3 H^2 \Lambda},
	\label{AI}
\end{eqnarray}
where we have used the scale factor $a\propto e^{H t}$ during the inflationary period.

\begin{figure}[t]
	\centering
	\includegraphics[scale=.8]{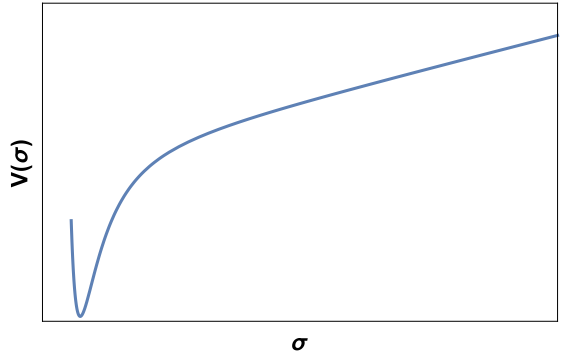}
	\caption{A conceptual shape of the dilaton potential is presented. At first, the background dilaton field rolls along the linear regime of its potential. The potential suddenly becomes steep at some point, and consequently the slow-roll motion of the dilaton is terminated. This is the point where the dilaton is stabilized, which corresponds to $N_D$.}
\label{fig:Dilaton potential}
\end{figure}

Let us now identify the parameter region which can realize our setup. First, the dilaton field needs to satisfy the slow-roll condition:
\dis{
\epsilon_\sigma\,=\,\frac{\Mp^2}{2}\bfrac{\partial_{\sigma}V}{V}^2\,\simeq\,  \frac{1}{18\Mp^2}\bfrac{\mu^3}{H^2}^2 \,\ll\, 1, \label{Model_SR}
}
as well as  the condition for the evolution rate of the dilaton-dependent gauge coupling:
\bea
n\,>\, 2+{\cal O}(0.1).
\eea
(Note that $\eta_\sigma = M_P^2 \partial_{\sigma}^2 V/V \,=\,0$ for the linear dilaton potential (\ref{dilaton-potential}).) Obviously these two conditions  can be satisfied with an appropriate choice of $\mu$ and $\Lambda$, e.g. 
\bea
\Lambda^2   \,\ll \, M_P^2, \quad \mu^3\,\,=\, {\cal O}(3H^2\Lambda)\, .
\eea

In order not to ruin the evolution of the inflaton field, the total variation of the dilaton potential over the rolling period  should be subdominant compared to the total energy density. This requires  
\dis{
\Delta V \,=\, \mu^3 (\sigma_{\rm in} - \sigma_D)
\,\simeq\, \frac{\mu^6 (N_T-N_D)}{3H^2} \,<\, 3M_P^2 H^2,
\label{sigma_const}
} 
where $\sigma_D\,\equiv\, \sigma(t_D)$ denotes the dilaton field value when the dilaton is stabilized, and  $N_T - N_D \,=\, (t_D-t_{\rm in})H$ denotes the number of e-folding that the inflationary universe has experienced  over the rolling period. This leads to an upper bound on the duration of the dilaton rolling as
\bea
N_T-N_D \,< \,  \bfrac{3H^2M_P}{\mu^3}^2 \,=\, \frac{1}{n^2}\left(\frac{\Mp}{\Lambda}\right)^2 \,\equiv\,  \Delta N_{\rm max}^{(1)}.
\label{BRBound0}
\eea

\begin{figure}[t]
	\centering
	\includegraphics[scale=.8]{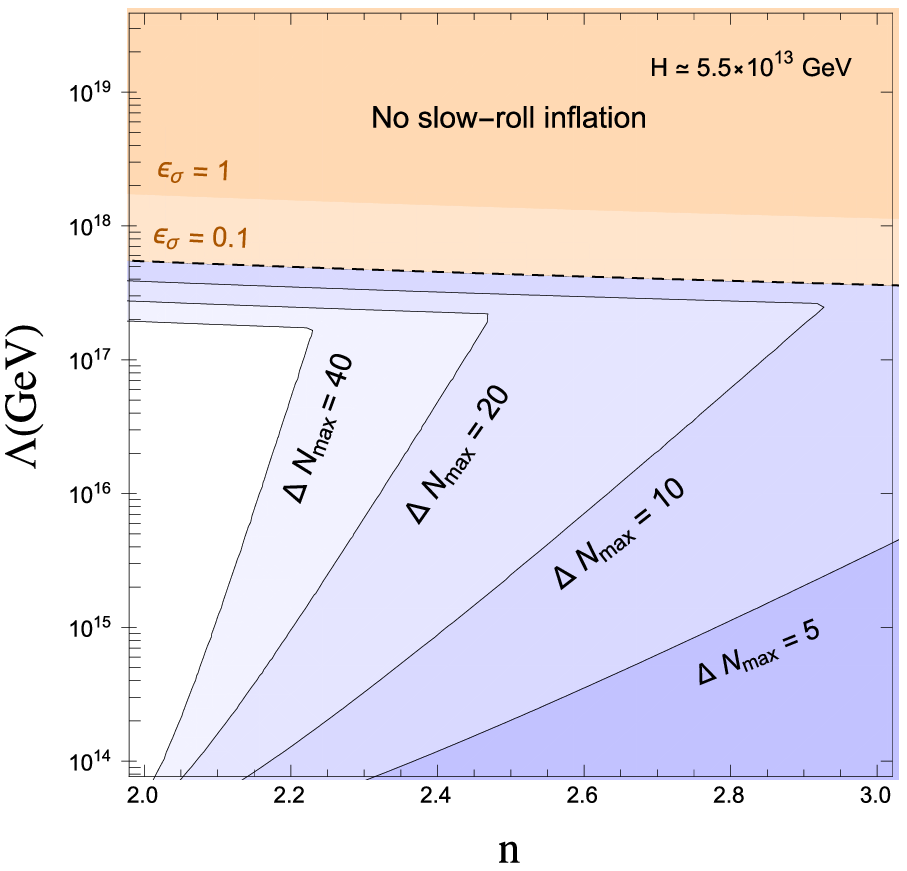}
	\hspace{.5cm}
	\includegraphics[scale=.8]{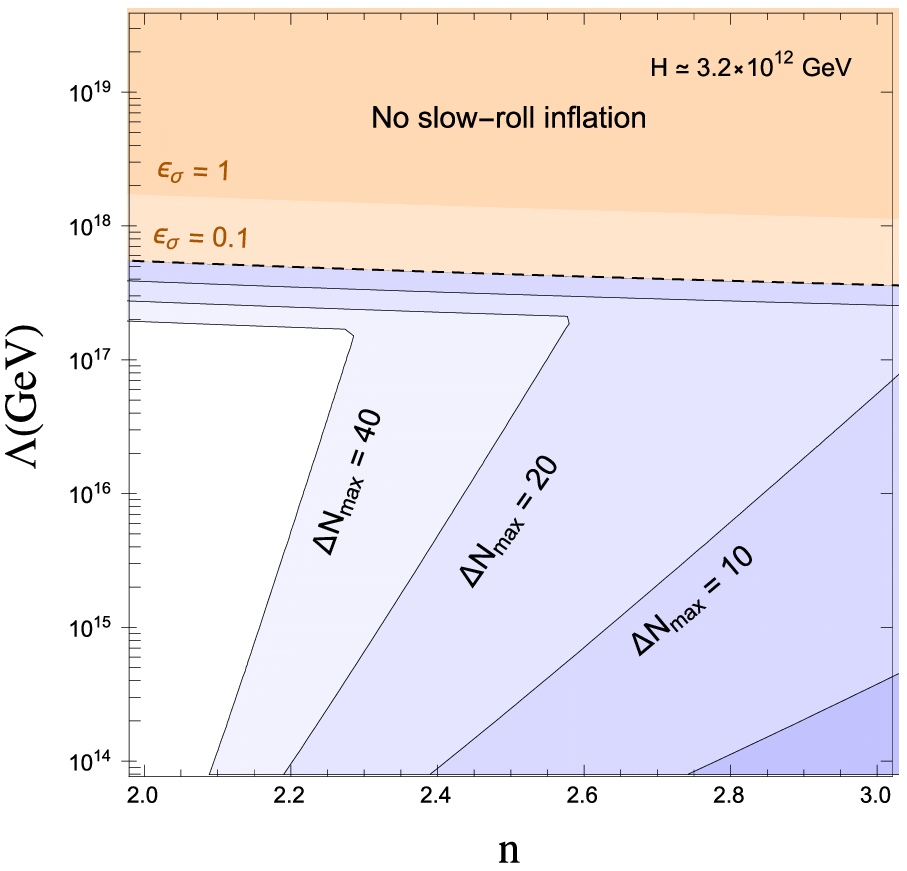}
	\caption
	{Constraint on the model parameters $n>2$ and $\Lambda$ for $H\simeq 5.5\times 10^{13} \gev$ (Left) and for $H\simeq3.2\times 10^{12} \gev$ (Right). In the orange region, the dilaton does not satisfy the slow roll condition~\eqref{Model_SR}. For $\Lambda \gtrsim 2\times10^{18} \gev$, the dilaton slow roll parameter exceeds the unity, $\epsilon_\sigma \gtrsim 1$, while for $\Lambda \gtrsim 5\times10^{17} \gev$, the dilaton slow roll parameter is $\epsilon_\sigma \gtrsim 0.1$, giving more stronger bound on this model. The contours in the blue area show the maximal duration of the dilaton slow-roll, allowed by the back-reaction constraints \eqref{BRBound0} and \eqref{BRBound}. }
	\label{BackR}
\end{figure}

So far, we have ignored  the effect of gauge fields on the evolution of the background dilaton field, which  would be justified only when 
\bea
\left| \partial_{\sigma} V  \right| \,\gg\,  	\left| \frac{\VEV{E^2}\partial_{\sigma}I}{I}\right|.
\label{BRC}
\eea
From \eqref{EMode}, we find
\bea
\frac{1}{2}\VEV{E^2}\, = \,\int \frac{d^3p}{(2\pi)^3} | {\cal E}(\tau,p) |^2
\,\simeq\, 
\frac{2^{2n-2}H^4}{\pi^3} \frac{\Gamma^2(n+1/2)}{(2n-4)} 
\left[e^{(2n-4) (N_T-N_D)} -1\right].
\eea
where the infrared divergence of the momentum integral is regulated by the cutoff $p_{\rm in}$ defined in \eqref{qin}.
Then the back-reaction constraint \eqref{BRC}  leads to an another bound on the duration of the dilaton rolling
(or the duration of the gauge field production):
\bea
N_T-N_D \,\lesssim\, \frac{1}{2n-4} 
\ln 
\left[ 
1 +  \frac{3 \pi^3}{2^{2n-2}} \frac{n (n-2)}{\Gamma^2(n+1/2)}  \left(\frac{\Lambda}{H}\right)^2 
\right]\,\equiv\, \Delta N_{\rm max}^{(2)}.
\label{BRBound}
\eea

There could be an additional constraint on the duration of the dilaton rolling from the requirement $\rho_{U(1)} \ll \rho_{\rm inflaton}$. However, since the $\rho_{U(1)}$ does not exceed the total variation of the dilaton energy density, the constraint \eqref{BRC}  automatically guarantees that $\rho_{U(1)} \ll \rho_{\rm inflaton}$. 

At this stage, we want to point out that a rapid oscillation of the dilaton during its stabilization process does not spoil the inflation. In this work, we have assumed that the variation of the dilaton energy density over the period of rolling dilaton is small enough compared to the inflation energy density. We can simply extrapolate this assumption to the regime where the dilaton is stabilized. The height of the steep dilaton potential would be naturally small compared to the scale of inflaton field. In such a case, the back-reaction by rapidly oscillating dilaton field can be safely neglected. See also \cite{Carney:2012pk} for how the sudden change in the dilaton potential can affect the primordial perturbations.

In figure \ref{BackR}, we depict the constraint on $n$ and $\Lambda$ for two different values of the inflationary Hubble scale $H\,\simeq\,(5.5\times 10^{13},\,3.2\times10^{12})\, \gev$. The contours show the maximal duration of the dilaton rolling, $\Delta N_{\rm max} = {\rm min} (\Delta N_{\rm max}^{(1)},\Delta N_{\rm max}^{(2)})$, allowed by the back-reaction constraints (\ref{BRBound0}) and (\ref{BRBound}). In the orange region above $\Lambda \gtrsim 2\times10^{18}\gev$, the background dilaton field $\sigma_0$ does not satisfy the slow-roll condition.

\begin{figure}[t]
	\centering
	\includegraphics[scale=.8]{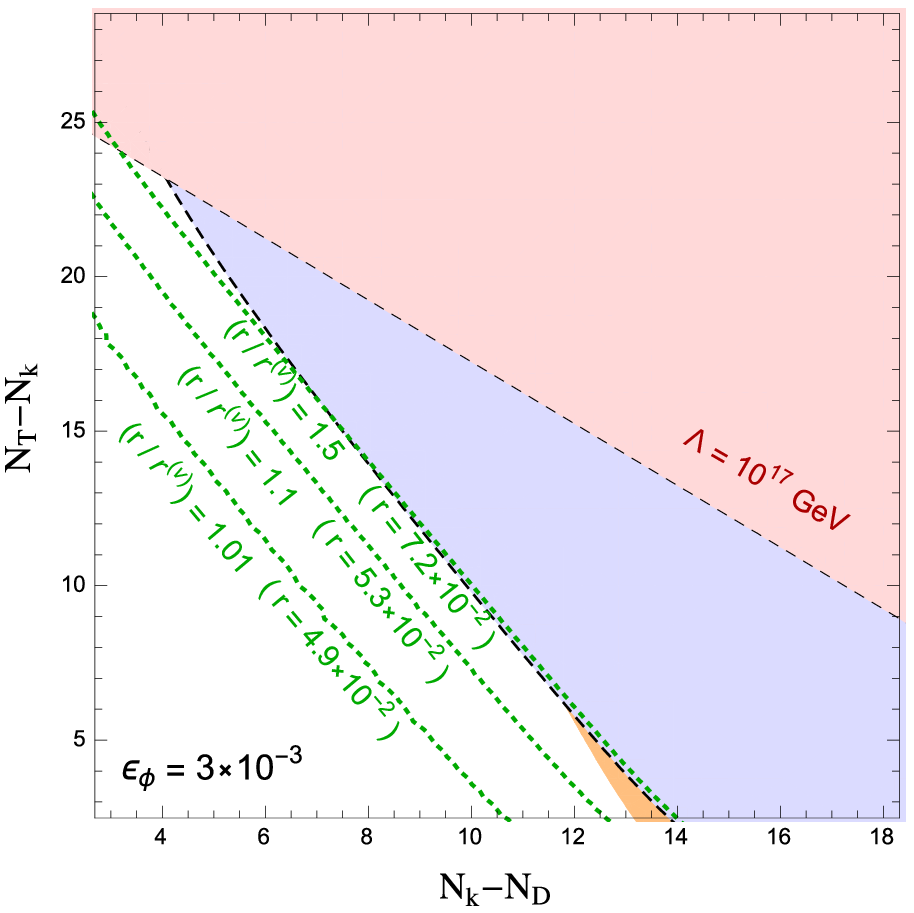}
	\hspace{.5cm}
	\includegraphics[scale=.8]{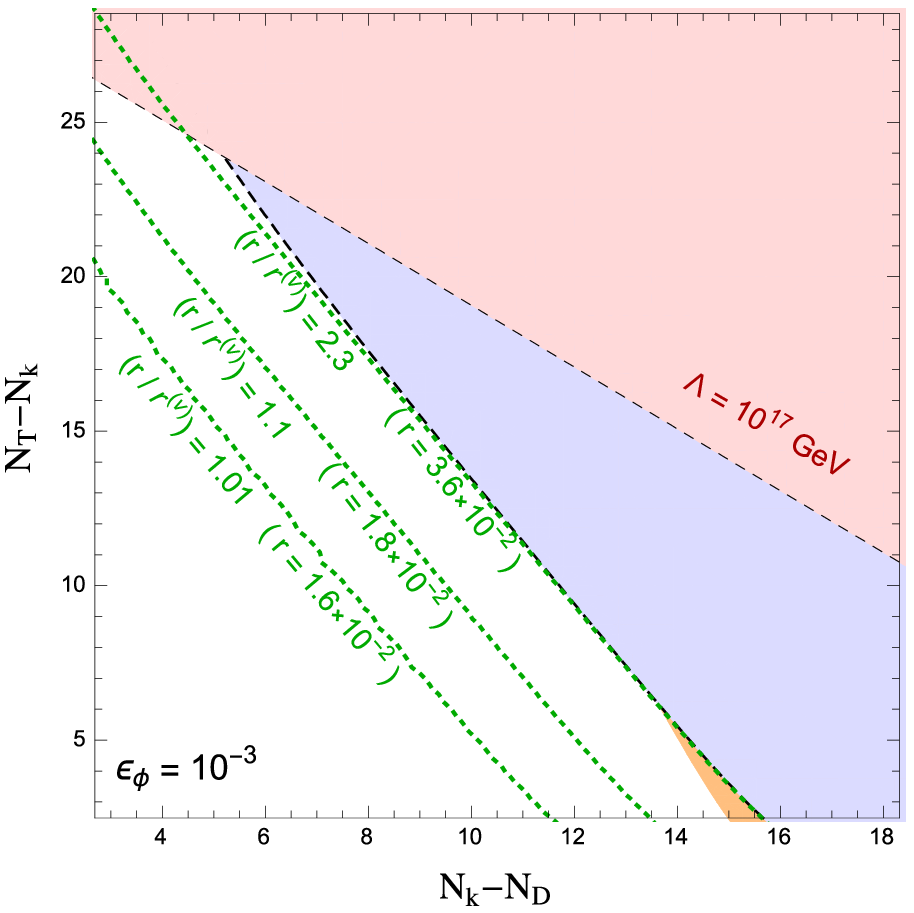}
	\includegraphics[scale=.8]{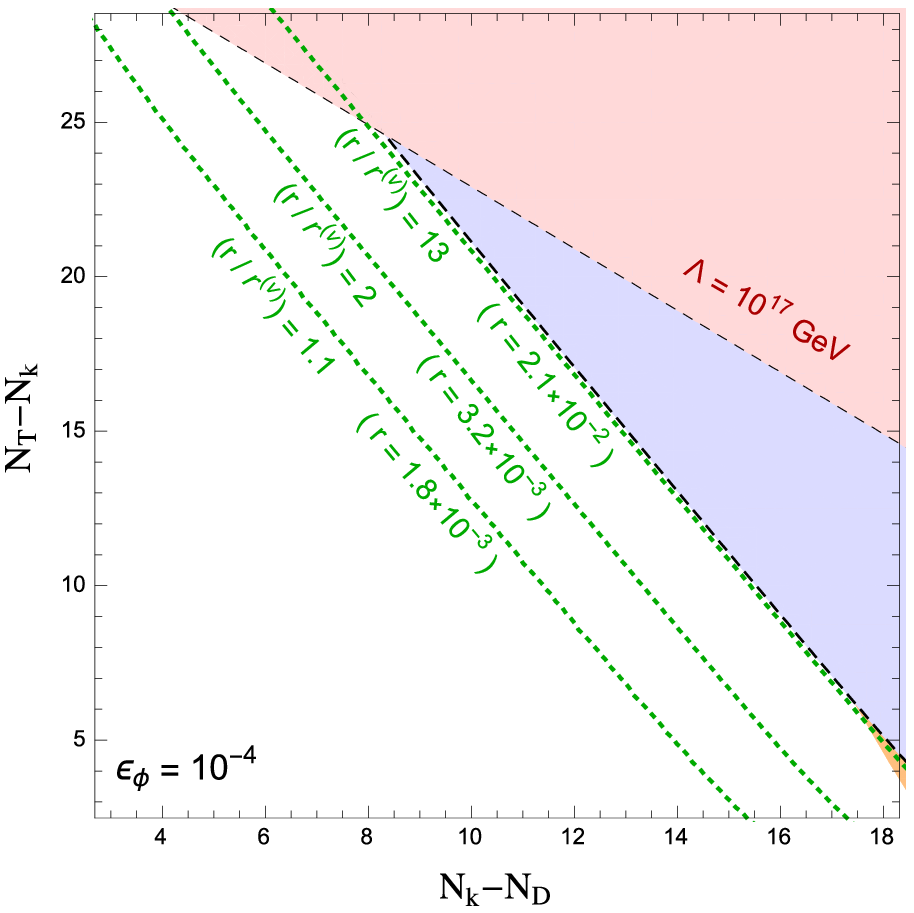}
	\hspace{.5cm}
	\includegraphics[scale=.8]{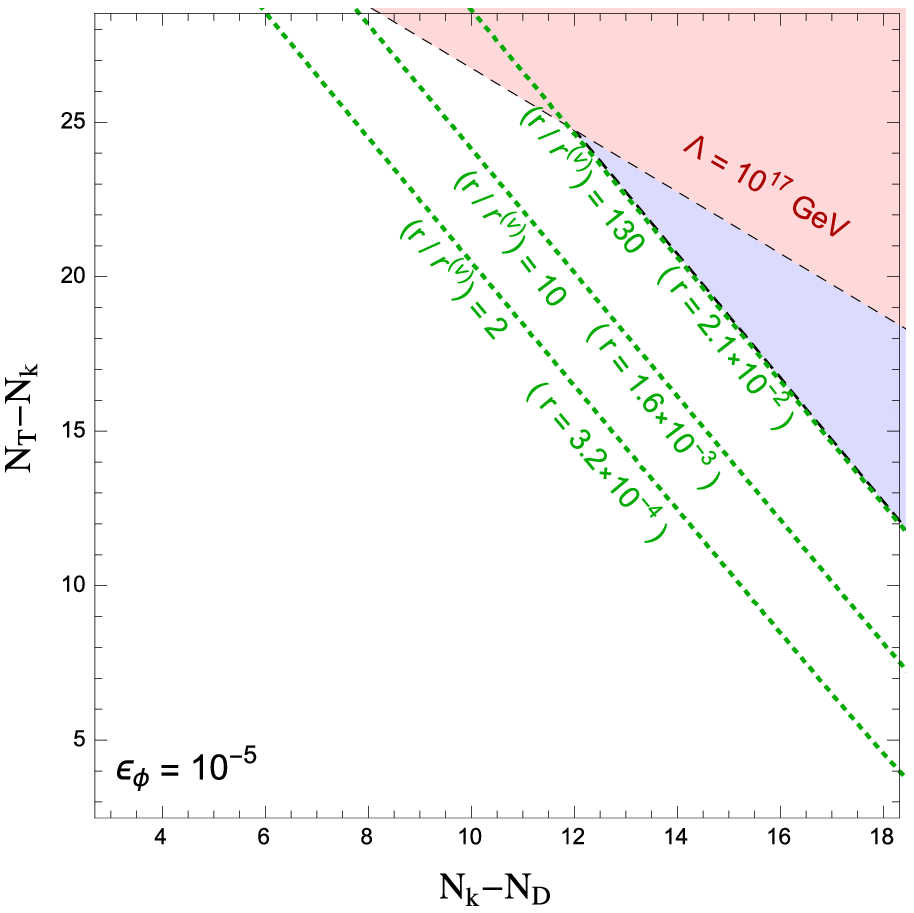}
	\caption{{The contour plots show $r$ and  $r/r^{(v)}$ for  $n=2.3$ and  $\epsilon_\phi=(3\times 10^{-3},\,10^{-3},\,10^{-4},\, 10^{-5})$. These figures are obtained by imposing the back-reaction constraints 
\eqref{BRBound0} and \eqref{BRBound} on figure~\ref{fig:r}. Note that the pink region excluded by the back-reaction constraints gives more stringent bound for large $N_T-N_k$ and small $N_k-N_D$ region, which is still allowed by spectral index constraint.}
}
\label{fig:rWb}
\end{figure}

In figure~\ref{fig:rWb}, we impose the back-reaction constraints \eqref{BRBound0} and \eqref{BRBound} on  figure~\ref{fig:r}, while assuming $\Lambda = 10^{17} \gev$. For this value of $\Lambda$, the pink region excluded by the back-reaction constraints are complementary to the constraint (\ref{const_scale_aniso}) from the scalar spectral index. It gives more stringent bound for large $N_T-N_k$ and small $N_k-N_D$ region, which is still allowed by the constraint from the spectral index.

\begin{figure}[t]
	\centering
	\includegraphics[scale=1]{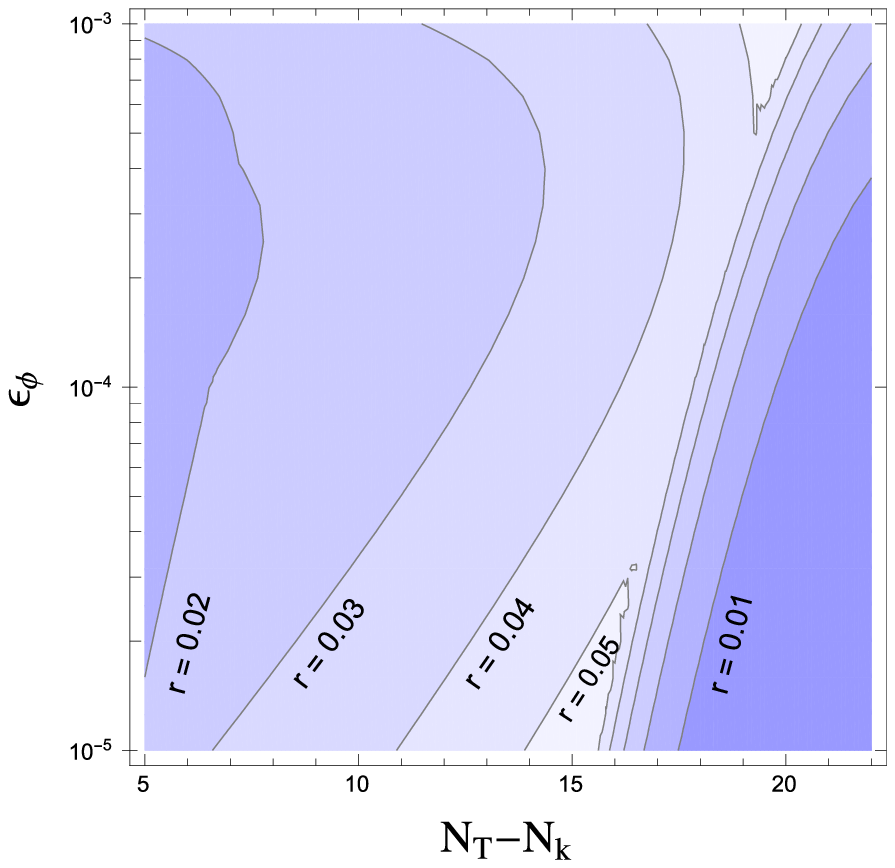}
	\caption
	{The contour plot for the maximal value of $r$ allowed by the observational constraints for $N_T -N_D = 25$.
	 For small value of $N_T- N_k$, 
	 the non-gaussianity and spectral index of the scalar power spectrum sourced by rolling dilaton provide the dominant constraints, while the back-reaction constraints dominate 
	 for large $N_T -N_k$.}
	\label{Rmax}
\end{figure}

In figure~\ref{Rmax}, we depict the maximal value of $r$ which can be achieved within our model, while satisfying all the available constraints, i.e. the spectral index constraint (\ref{const_scale_aniso}), the non-gaussianity constraint (\ref{non-gaussianity-constraint}), and the back-reaction constraints (\ref{BRBound0}) and (\ref{BRBound}) for $N_T- N_D=25$ and $\Lambda=10^{17}$ GeV. For this, we assumed  the relation $\epsilon_\phi\simeq 0.01 \,(H/10^{14}\, {\rm GeV})^2$ which would be valid in generic single field inflation scenario involving a spectator dilaton. For small  $N_T- N_k$, the spectral index and non-gaussianity of the scalar perturbation  sourced by rolling dilaton provide the dominant constraints on the possible value of $r$. On the other hand, for given values of $\epsilon_\phi$ and $N_T-N_D$,  a larger $N_T- N_k$ gives a smaller scalar power spectrum ${\cal P}^{(s)}_{\mathcal R}$ sourced by rolling dilaton (see (\ref{PRsN})), and as a result the constraints from the spectral index and non-gaussianity becomes less important, and $r$ is limited dominantly by the back-reaction constraints. Note that the maximal value of $r$ is not so sensitive to the value of $\epsilon_\phi$. This can be anticipated from the bound (\ref{UB}) which can be nearly saturated in most cases. Here we can see that our model can give $r\gtrsim 10^{-2}$   which is large enough to be  probed in the near future \cite{Creminelli:2015oda}, even when the inflation scale is relatively low to yield $\epsilon_\phi \ll 10^{-3}$.

\section{Conclusion} \label{conclusion}
Fundamental theories for physics beyond the standard model of particle physics often involve a dilaton field $\sigma$ which results in a
dilaton-dependent gauge coupling $g(\sigma)=1/I(\sigma)$. Under a reasonable assumption on the time evolution
of $\sigma$ during the inflation epoch, the dilaton coupling to gauge fields can evolve as
  $I\propto a^{-n}$, where $n$ is approximately a constant.  
If this evolution rate $n$  is as large as $n >2$, the energy density of  
the gauge fields produced by rolling dilaton  can grow in the superhorizon regime, and thereby significantly affect the primordial perturbations.

In this paper, we have examined the possibility to have an observably large tensor-to-scalar ratio with the tensor perturbation generated by the gauge fields produced by rolling dilaton. As the dilaton-induced gauge fields generate a scalar perturbation  which is strongly  scale-dependent and non-gaussian, this scheme is severely constrained by the observed approximately scale-invariant and gaussian scalar perturbation. Yet, we find that $r$ can be significantly  enhanced relative to the standard result $r=16\epsilon$ while satisfying the observational constraints, if the dilaton is stabilized before the end of inflation, but after the horizon exit of the CMB scale.

Imposing the observational constraints on the scalar perturbation sourced by rolling dilaton, we find that for the inflaton slow roll parameter $\epsilon_\phi  \gtrsim 10^{-3}$, the  tensor-to-scalar ratio  can be modified  only by a factor of ${\cal O}(1)$ compared to the standard result. However, for smaller $\epsilon_\phi$, which corresponds to lower inflation energy scale, $r$ can be enhanced by a much larger factor. As a consequence, the tensor perturbation sourced by rolling dilaton can give rise to an observably large $r\gtrsim  10^{-2}$ even when the inflaton slow roll parameter $\epsilon_\phi\ll 10^{-3}$. Contrary to the one from the vacuum fluctuation of the metric, the tensor perturbation sourced by rolling dilaton generically has a strongly red-tilted scale dependence, i.e. $n_t=-2(n-2)$ whose magnitude can be of order unity. As discussed in Section~\ref{model}, our scenario can be realized within the framework of a simple model in a self consistent manner.

\section*{Acknowledgments}
We thank Shinji Mukoyama, Marco Peloso, Gary Shiu and J. Yokoyama for useful discussions and comments. This work was supported by IBS under the project code, IBS-R018-D1. C.S.S. are supported in part by DOE grants DOE-SC0010008, DOE-ARRA-SC0003883, and DOE-DE-SC0007897.

\appendix

\section{Equations of motion of perturbations} \label{SourceDerivation}

In this appendix, we derive the equations of motion (\ref{IPEOM}) and (\ref{DPEOM}) for the inflaton and dilaton perturbations. For the action functional \eqref{Model} and the spacetime metric \eqref{metric} in the flat gauge, the equations of motion for the inflaton and dilaton fields are given by
\bea
-\frac{1}{\sqrt{-g}} \partial_\mu ( \sqrt{-g} g^{\mu\nu} \partial_\nu \phi ) 
+ \partial_\phi V &=&0, \nonumber
\\
-\frac{1}{\sqrt{-g}} \partial_\mu ( \sqrt{-g} g^{\mu\nu} \partial_\nu \sigma )
 + \partial_\sigma V &=& -\frac{1}{2}I\partial_\sigma I  F_{\mu\nu}F^{\mu\nu}.
\nonumber \eea
Splitting the scalar fields into the backgrounds and the fluctuations, we obtain 
\begin{eqnarray}
	\phi_0'' + 2 {\cal H} \phi_0' + a^2 \partial_{\phi_0}V(\phi_0) &=& 0, \nonumber
	\\
	\sigma_0'' + 2 {\cal H} \sigma_0' + a^2 \partial_{\sigma_0}V(\sigma_0) &=& 
	a^2 \frac{\partial_{\sigma_0}I(\sigma_0)}{I(\sigma_0)} \VEV{E^2 - B^2}, \nonumber
\end{eqnarray}
and 
\bea
\delta\phi '' + 2 {\cal H} \delta\phi' 
+ a^2\left( \partial_{\phi_0}^2V + \frac{k^2}{a^2} \right) \delta\phi
+ k^2 \phi_0' B - \phi_0' \Phi' + 2 a^2 \Phi \partial_{\phi_0}V &=& 0, \nonumber
\\
\delta\sigma '' + 2 {\cal H} \delta\sigma' 
+ a^2\left( \partial_{\sigma_0}^2V + \frac{k^2}{a^2} \right) \delta\sigma
+ k^2 \sigma_0' B - \sigma_0' \Phi' + 2 a^2 \Phi\partial_{\sigma_0}V &=& S_2, \nonumber
\eea
where
\bea
S_2(\tau,\bfk) = \frac{a^2 I_{,\sigma}}{I} \left[ (E^2-B^2) - \VEV{E^2-B^2} \right](\tau,\bfk). \nonumber
\eea

Similary, from the Einstein equation, one finds the following equations of motion 
for the metric perturbations in the flat gauge: 
\begin{eqnarray}
	3H^2 \Phi - \frac{H}{a} k^2 B &=&  
	- \frac{\delta \rho}{2\Mp^2}, \nonumber
	\\
	H \Phi &=&  - \frac{\delta q}{2 \Mp^2}, \nonumber
	\\
	H \dot{\Phi} + (3 H^2 + 2 \dot{H} ) \Phi &=& 
	\frac{1}{2\Mp^2} (\delta p - \frac{2}{3} \Pi), \nonumber
	\\
	-\frac{k^2}{a^2} (\Phi + a \dot{B} + 2 a H B) 
	&=& \frac{\Pi}{\Mp^2}, \nonumber
\end{eqnarray}
where $\delta p$, $\delta q$, and $\Pi$ are defined through the fluctuation of the energy momentum tensor as
\bea
\delta T^t_{\; i} &=& \partial_i \delta q,\nonumber \\
\delta T^i_{\; k} &=& \delta p \; \delta^i_{\; j} + \Pi^i_{\; j}, \nonumber \eea
for the anisotropy tensor $\Pi^i_{\; j}$ decomposed as 
\bea
\Pi^i_{\; j} = 
\left( \frac{1}{3} \delta^i_j - \frac{\partial_i \partial_j}{\nabla^2} \right) \Pi
+ \frac{1}{2} (\Pi^{(v),i}_{\quad \;\; j} + \Pi^{(v)i}_{\quad \;\; ,j}) + \Pi^{(t)i}_{\quad \;\; j}.\nonumber 
\eea

Then, assuming the slow-roll motion of the involved scalar fields, we obtain
\bea
\delta\phi '' + 2 {\cal H} \delta\phi' 
+ a^2\left( \partial_\phi^2V + \frac{k^2}{a^2} \right) \delta\phi
&=& 
-\frac{a^2 \dot \phi_0}{2 \Mp^2 H} 
\left[ (\delta \rho - \delta p) + \frac{2}{3} \Pi \right] 
- \epsilon \frac{a^2  \dot \phi_0}{\Mp^2 H} H\delta q,
\nonumber\\
\delta\sigma '' + 2 {\cal H} \delta\sigma' 
+ a^2\left( \partial_{\sigma}^2V + \frac{k^2}{a^2} \right) \delta\sigma
&=& S_2 -\frac{a^2 \dot \sigma_0}{2 \Mp^2 H} 
\left[ (\delta \rho - \delta p) + \frac{2}{3} \Pi \right] 
- \epsilon \frac{a^2  \dot \sigma_0}{\Mp^2 H} H\delta q.
\nonumber
\eea
where $\epsilon \equiv -\dot{H}/H^2$. Under the slow-roll assumption,  the last term in the right hand side of these
equations of motion can be neglected. In our setup, the energy momentum tensor is determined by
the inflation, dilaton, and gauge fields, which leads to the relations
\bea
&&\left[ (\delta \rho - \delta p) + \frac{2}{3} \Pi \right]_{\rm inflaton} 
\simeq -6 H \dot{\phi}_0 \delta\phi, \nonumber
\\
&&\left[ (\delta \rho - \delta p) + \frac{2}{3} \Pi \right]_\dilaton 
\simeq -6 H \dot{\sigma}_0 \delta\sigma, \nonumber
\\
&&\left[ (\delta \rho - \delta p) + \frac{2}{3} \Pi \right]_{\rm gauge} 
\simeq \hat{k}_i \hat{k}_j (E_i E_j + B_i B_j)(\tau,\bfk). \nonumber
\eea
Putting these into the equations of motion of the scalar field fluctuations, we finally find
\begin{eqnarray}
	\delta \phi '' + 2 {\cal H} \delta\phi' + k^2 \delta \phi
	+ a^2 \left( \partial_\phi^2V - 3 \frac{\dot{\phi}^2_0}{\Mp^2} \right) 
	\delta \phi
	- 3 a^2 \frac{\dot{\sigma}_0 \dot{\phi}_0}{\Mp^2} \delta \sigma
	&=&  S_1 , \nonumber
	\\
	\delta \sigma '' + 2 {\cal H} \delta\sigma' + k^2 \delta \sigma
	+ a^2 
	\left( \partial_\sigma^2V - 3 \frac{\dot{\sigma}^2_0}{\Mp^2} \right) 
	\delta \sigma
	- 3 a^2 \frac{\dot{\sigma}_0 \dot{\phi}_0}{\Mp^2} \delta \phi
	&=&  S_2 + S_3, \nonumber
\end{eqnarray}
where the three source terms are given by
\bea
S_1(\tau,\bfk) &=& -\frac{a^2 \dot{\phi}_0}{2\Mp^2 H} 
\int \frac{d^3\bfp}{(2\pi)^{3/2}} \frac{(k_i-p_i) p_j}{k^2}
\left[
\widehat{E}_i(\tau,{\bf p}) \widehat{E}_j(\tau,{\bfk - \bf p}) + 
\widehat{B}_i(\tau,{\bf p}) \widehat{B}_j(\tau,{\bfk - \bf p})
\right],
\nonumber\\
S_2(\tau,\bfk) &=& a^2\frac{I_{,\sigma}}{I}
\int \frac{d^3\bfp}{(2\pi)^{3/2}} 
\left[
\widehat{E}_i(\tau,{\bf p}) \widehat{E}_i(\tau,{\bfk - \bf p}) + 
\widehat{B}_i(\tau,{\bf p}) \widehat{B}_i(\tau,{\bfk - \bf p})
\right],
\nonumber\\
S_3(\tau,\bfk) &=& -\frac{a^2 \dot{\sigma}_0}{2\Mp^2 H} 
\int \frac{d^3\bfp}{(2\pi)^{3/2}} \frac{(k_i-p_i) p_j}{k^2}
\left[
\widehat{E}_i(\tau,{\bf p}) \widehat{E}_j(\tau,{\bfk - \bf p}) + 
\widehat{B}_i(\tau,{\bf p}) \widehat{B}_j(\tau,{\bfk - \bf p})
\right].
\nonumber
\eea

\section{Green function}\label{Green}
Here we compute the Green function up to first order in the slow-roll parameters. For the differential equation
\begin{eqnarray}
	\left[ 
	\partial_\tau^2 + 
	\left( k^2 - \frac{2 - \Delta_{\pm}}{\tau^2} \right) \right]
	G_k (\tau,\tau'; \Delta_{\pm}) = \delta(\tau-\tau'),
	\label{GreenDE}
\end{eqnarray}
the solution is given by
\begin{eqnarray}
	G_k(\tau,\tau';\Delta_{\pm}) = i \Theta(\tau-\tau')
	\left[ 
		Q_k(\tau;\Delta_{\pm}) Q^*_k(\tau';\Delta_{\pm})
		-
		Q^*_k(\tau;\Delta_{\pm}) Q_k(\tau';\Delta_{\pm})
	\right],\nonumber
\end{eqnarray}
where $Q_k(\tau)$ denotes the homogeneous solution of (\ref{GreenDE}).
One then finds 
\begin{eqnarray}
	Q_k(\tau; \Delta_\pm) = \frac{1}{\sqrt{2k}} \sqrt{\frac{-k\tau\pi}{2}} 
	H^{(1)}_{\nu_\pm}(-k\tau),\nonumber
\end{eqnarray}
which yields  
\begin{eqnarray}
	G_k(\tau,\tau'; \Delta_{\pm}) =
	\frac{i\pi}{4} \sqrt{\tau\tau'} \Theta(\tau-\tau')
	\left[  
		H^{(1)}_{\nu_\pm} (-k\tau) H^{(1)*}_{\nu_\pm} (-k\tau') 
		- H^{(1)*}_{\nu_\pm} (-k\tau) H^{(1)}_{\nu_\pm} (-k\tau') 
	\right],\nonumber
\end{eqnarray}
where the order of the Hankel function is given by
$$
	\nu_\pm \,=\,  \frac{3}{2} \sqrt{ 1 - \frac{4\Delta_{\pm}^2}{9H^2}}
	\,\simeq\, \frac{3}{2} - \frac{\Delta_{\pm}}{3}.
$$
Here we  are interested in the perturbation at superhorizon scale $|k\tau| \ll1$,
produced by gauge fields at another  superhorizon scale  $|k\tau'| \ll1$. Then, using the asymptotic form of the Hankel function $H_\nu^{(1)}(x)$ in the limit $x\ll1$,
\bea
H^{(1)}_\nu (x) \simeq 
-\frac{i \Gamma(\nu)}{\pi} \left( \frac{2}{x}\right)^\nu
+ \frac{1}{\Gamma(\nu+1)} \left(\frac{x}{2} \right)^\nu 
- \frac{i\Gamma(-\nu)}{\pi} \cos \nu \pi \left(\frac{x}{2}\right)^\nu,\nonumber
\eea
we find 
\begin{eqnarray}
	G_k(\tau,\tau';\Delta_{\pm}) &\simeq&
	\Theta(\tau-\tau')
	\frac{\sqrt{\tau\tau'}}{2}
	 \frac{\Gamma(\nu_\pm)}{\Gamma(\nu_\pm+1)} 
	\left[
	\left( \frac{\tau'}{\tau} \right)^{\nu_{\pm}} -
	\left( \frac{\tau}{\tau'} \right)^{\nu_{\pm}}
	\right],\nonumber
\end{eqnarray}
which results in
\bea
	&&G_k(\tau,\tau';\Delta_{+}) - G_k(\tau,\tau';\Delta_{-})
	\nonumber \\
	&&\qquad\qquad\simeq  \Theta(\tau-\tau')
	\left(\frac{\Delta_- - \Delta_+}{9} \right)
	\left( \frac{\tau'^2}{\tau} \right)
	\left[ \left(\ln\frac{\tau'}{\tau} - \frac{2}{3}  \right) - \frac{\tau^3}{\tau'^3} 
	\left( \ln \frac{\tau}{\tau'} - \frac{2}{3}\right)  \right].
	\label{GDifference}
\eea


\end{document}